\documentclass[preprint]{aastex631}  %
\received{2023 December 16}
\revised{---}
\accepted{---}

\submitjournal{ApJ}

\shorttitle{Forming planets: faint accretion tracers}  %
\shortauthors{G.-D.\ Marleau}
\usepackage{etoolbox}

\makeatletter

\patchcmd{\NAT@citex}
  {\@citea\NAT@hyper@{%
     \NAT@nmfmt{\NAT@nm}%
     \hyper@natlinkbreak{\NAT@aysep\NAT@spacechar}{\@citeb\@extra@b@citeb}%
     \NAT@date}}
  {\@citea\NAT@nmfmt{\NAT@nm}%
   \NAT@aysep\NAT@spacechar\NAT@hyper@{\NAT@date}}{}{}

\patchcmd{\NAT@citex}
  {\@citea\NAT@hyper@{%
     \NAT@nmfmt{\NAT@nm}%
     \hyper@natlinkbreak{\NAT@spacechar\NAT@@open\if*#1*\else#1\NAT@spacechar\fi}%
       {\@citeb\@extra@b@citeb}%
     \NAT@date}}
  {\@citea\NAT@nmfmt{\NAT@nm}%
   \NAT@spacechar\NAT@@open\if*#1*\else#1\NAT@spacechar\fi\NAT@hyper@{\NAT@date}}
  {}{}

\makeatother

\def\degr{{\mbox{\textdegree}}\xspace}

\usepackage{accsupp}
\providecommand*{\napprox}{%
  \BeginAccSupp{method=hex,unicode,ActualText=2249}%
  \not\approx
  \EndAccSupp{}%
}

\makeatletter
\let\jnl@style=\rm
\def\ref@jnl#1{{\jnl@style#1}}

\def\aj{\ref@jnl{AJ}}                   %
\def\actaa{\ref@jnl{Acta Astron.}}      %
\def\araa{\ref@jnl{ARA\&A}}             %
\def\apj{\ref@jnl{ApJ}}                 %
\def\apjl{\ref@jnl{ApJ}}                %
\def\apjs{\ref@jnl{ApJS}}               %
\def\ao{\ref@jnl{Appl.~Opt.}}           %
\def\apss{\ref@jnl{Ap\&SS}}             %
\def\aap{\ref@jnl{A\&A}}                %
\def\aapr{\ref@jnl{A\&A~Rev.}}          %
\def\aaps{\ref@jnl{A\&AS}}              %
\def\azh{\ref@jnl{AZh}}                 %
\def\baas{\ref@jnl{BAAS}}               %
\def\bac{\ref@jnl{Bull. astr. Inst. Czechosl.}}
\def\caa{\ref@jnl{Chinese Astron. Astrophys.}}
\def\cjaa{\ref@jnl{Chinese J. Astron. Astrophys.}}
\def\icarus{\ref@jnl{Icarus}}           %
\def\jcap{\ref@jnl{J. Cosmology Astropart. Phys.}}
\def\jrasc{\ref@jnl{JRASC}}             %
\def\memras{\ref@jnl{MmRAS}}            %
\def\mnras{\ref@jnl{MNRAS}}             %
\def\na{\ref@jnl{New A}}                %
\def\nar{\ref@jnl{New A Rev.}}          %
\def\pra{\ref@jnl{Phys.~Rev.~A}}        %
\def\prb{\ref@jnl{Phys.~Rev.~B}}        %
\def\prc{\ref@jnl{Phys.~Rev.~C}}        %
\def\prd{\ref@jnl{Phys.~Rev.~D}}        %
\def\pre{\ref@jnl{Phys.~Rev.~E}}        %
\def\prl{\ref@jnl{Phys.~Rev.~Lett.}}    %
\def\pasa{\ref@jnl{PASA}}               %
\def\pasp{\ref@jnl{PASP}}               %
\def\pasj{\ref@jnl{PASJ}}               %
\def\rmxaa{\ref@jnl{Rev. Mexicana Astron. Astrofis.}}%
\def\rnaas{\ref@jnl{RNAAS}}             %
\def\qjras{\ref@jnl{QJRAS}}             %
\def\skytel{\ref@jnl{S\&T}}             %
\def\solphys{\ref@jnl{Sol.~Phys.}}      %
\def\sovast{\ref@jnl{Soviet~Ast.}}      %
\def\ssr{\ref@jnl{Space~Sci.~Rev.}}     %
\def\zap{\ref@jnl{ZAp}}                 %
\def\nat{\ref@jnl{Nature}}              %
\def\iaucirc{\ref@jnl{IAU~Circ.}}       %
\def\aplett{\ref@jnl{Astrophys.~Lett.}} %
\def\apspr{\ref@jnl{Astrophys.~Space~Phys.~Res.}}
\def\bain{\ref@jnl{Bull.~Astron.~Inst.~Netherlands}} 
\def\fcp{\ref@jnl{Fund.~Cosmic~Phys.}}  %
\def\gca{\ref@jnl{Geochim.~Cosmochim.~Acta}}   %
\def\grl{\ref@jnl{Geophys.~Res.~Lett.}} %
\def\jcp{\ref@jnl{J.~Chem.~Phys.}}      %
\def\jgr{\ref@jnl{J.~Geophys.~Res.}}    %
\def\jqsrt{\ref@jnl{J.~Quant.~Spec.~Radiat.~Transf.}}
\def\memsai{\ref@jnl{Mem.~Soc.~Astron.~Italiana}}
\def\nphysa{\ref@jnl{Nucl.~Phys.~A}}   %
\def\physrep{\ref@jnl{Phys.~Rep.}}   %
\def\physscr{\ref@jnl{Phys.~Scr}}   %
\def\planss{\ref@jnl{Planet.~Space~Sci.}}   %
\def\procspie{\ref@jnl{Proc.~SPIE}}   %

\def\ptp{\ref@jnl{Prog.~Th.~Phys.}}   %
\def\natas{\ref@jnl{NatAs}}           %
\def\amjm{\ref@jnl{AmJM}}             %

\makeatother

\defcitealias{goli04}{G04}
\defcitealias{marl07}{M07}
\defcitealias{scvh}{SCvH}
\defcitealias{bl94}{BL94}
\defcitealias{mc14}{MC14}
\defcitealias{ensman94}{E94}
\defcitealias{lever81}{LP81}

\defcitealias{m22Schock}{MKBM23}

\defcitealias{boden13}{B13}
\defcitealias{emsen21a}{NGPPS}

\defcitealias{ab22}{AB22}
\def\ab{\citetalias{ab22}\xspace}
\def\AB{\citepalias{ab22}\xspace}

\defcitealias{ab25}{AB25}
\def\abb{\citetalias{ab25}\xspace}

\usepackage{xspace}

\usepackage{amsmath,amssymb,listings,upgreek,nicefrac}
\usepackage{newtxtext,newtxmath}
\usepackage{grffile}
\usepackage{xcolor,xspace}
\usepackage[normalem]{ulem}  %

\usepackage[T1]{fontenc}
\usepackage{textcomp}     %

\DeclareMathOperator\erf{erf}

\def\e{\textrm{e}}                                 %
\def\mH{m_\textrm{H}}                              %
\def\kB{k_\textrm{B}}                              %
\def\sigSB{\ensuremath{\sigma_{\textrm{SB}}}\xspace}        %
\def\aST{\ensuremath{a_{\textrm{r}}}\xspace}       %
\def\MJ{\ensuremath{M_{\textrm{J}}}\xspace}        %
\def\RJ{\ensuremath{R_{\textrm{J}}}\xspace}        %
\def\ME{\ensuremath{M_{\textrm{E}}}\xspace}        %
\def\MSonne{\ensuremath{M_\odot}\xspace}           %
\def\LSonne{\ensuremath{L_\odot}\xspace}           %

\def\Ha{H$\alpha$\xspace}                               %
\def\Paa{Pa$\alpha$\xspace}                             %
\def\Pab{Pa$\beta$\xspace}                              %
\def\Pag{Pa$\gamma$\xspace}                             %
\def\Bra{Br$\alpha$\xspace}                             %
\def\Brg{Br$\gamma$\xspace}                             %
\def\PDS{PDS\,70\xspace}                                  %
\def\PDSb{PDS\,70\,b\xspace}                              %
\def\PDSc{PDS\,70\,c\xspace}                              %
\def\PDSbc{PDS\,70\,b and~c\xspace}                       %
\def\Dlrmb{Delorme\,1\,(AB)b\xspace}                      %

\def\MPkt{\ensuremath{\dot{M}}\xspace}                               %
\def\MPkthineinABT{\ensuremath{\dot{M}_{\textrm{in}}}\xspace}            %
\def\MPktnettoHill{\ensuremath{\dot{M}_{\textrm{Hill,\,net}}}\xspace}       %
\def\MPktvekst{\ensuremath{\dot{M}_{\textrm{growth}}}\xspace}    %
\def\MPktPopsynth{\ensuremath{\dot{M}_{\textrm{pop synth}}}\xspace}    %
\def\MPktPdir{\ensuremath{\dot{M}_{\textrm{p,\,direct}}}\xspace}      %
\def\MPktPdirABT{\ensuremath{\dot{M}_{\textrm{p,\,dir,\,ABT}}}\xspace}      %
\def\fdir{\ensuremath{f_{\textrm{dir}}}\xspace}                    %

\def\LPlObfl{\ensuremath{L_{\textrm{plnt~surf}}}\xspace}   %
\def\LzpSch{\ensuremath{L_{\textrm{CPD~surf}}}\xspace}     %

\def\fLuecke{\ensuremath{f_{\textrm{gap}}}\xspace}   %
\def\FBoden{\ensuremath{F_{\textrm{\citetalias{boden13}}}}\xspace}      %

\def\Derf{\ensuremath{\Delta\!\erf}\xspace}      %

\def\MP{\ensuremath{M_{\textrm{p}}}\xspace}        %
\def\RP{\ensuremath{R_{\textrm{p}}}\xspace}        %
\def\MzpSch{\ensuremath{M_{\textrm{CPD}}}\xspace}           %
\newcommand{\Lacc}{\ensuremath{{L_{\textrm{acc}}}}\xspace}  %
\def\LAkk{\Lacc}
\newcommand{\Tacc}{\ensuremath{{T_{\textrm{acc}}}}\xspace}  %
\def\LHa{\ensuremath{L_{\textrm{H}\,\alpha}}\xspace}    %
\def\FHa{\ensuremath{F_{\textrm{H}\,\alpha}}\xspace}             %
\def\LPab{\ensuremath{L_{\textrm{Pa}\,\beta}}\xspace}    %
\def\FPab{\ensuremath{F_{\textrm{Pa}\,\beta}}\xspace}    %
\def\LLinie{\ensuremath{L_{\textrm{line}}}\xspace}             %
\newcommand{\RH}{\ensuremath{{R_{\textrm{Hill}}}}\xspace}      %
\def\RHill{\RH}
\newcommand{\RB}{\ensuremath{{R_{\textrm{Bondi}}}}\xspace}     %
\newcommand{\HPPPD}{\ensuremath{H_{P,\,\textrm{PPD}}}\xspace}  %
\newcommand{\HPzpSch}{\ensuremath{H_{P,\,\textrm{CPD}}}\xspace} %
\newcommand{\Tint}{\ensuremath{T_{\textrm{int}}}\xspace}        %
\newcommand{\Teff}{\ensuremath{T_{\textrm{eff}}}\xspace}        %
\def\cs{\ensuremath{c_{\textrm{s}}}\xspace}        %
\def\kappaStbint{\ensuremath{\kappa_{\bullet}}\xspace}  %
\def\fpg{\ensuremath{f_{\textrm{d/g}}}\xspace}
\newcommand{\mPunkt}{\dot{m}}

\newcommand{\vVorSch}{\ensuremath{v_{\textrm{pre}}}\xspace}
\newcommand{\rhoVorSch}{\ensuremath{\rho_{\textrm{pre}}}\xspace}
\newcommand{\nVorSch}{\ensuremath{n_{\textrm{pre}}}\xspace}

\newcommand{\MStern}{\ensuremath{{M_{\star}}}\xspace}
\newcommand{\RStern}{\ensuremath{{R_{\star}}}\xspace}
\newcommand{\LStern}{\ensuremath{{L_{\star}}}\xspace}

\newcommand{\qth}{\ensuremath{{q_{\textrm{th}}}}\xspace}

\newcommand{\rmax}{\ensuremath{{r_{\textrm{max}}}}\xspace}

\newcommand{\Rzent}{\ensuremath{{R_{\textrm{cent}}}}\xspace}

\newcommand{\vFfinfty}{\ensuremath{{v_{\textrm{ff},\,\infty}}}\xspace}

\newcommand{\vkrit}{\ensuremath{{v_{\textrm{crit}}}}\xspace}

\newcommand{\rhoMitt}{\ensuremath{{\rho_{\textrm{mid}}}}\xspace}

\newcommand{\TAkk}{\Tacc}   %

\newcommand{\munaufp}{\ensuremath{\mu_{0\rightarrow\textrm{p}}}\xspace}
\newcommand{\thnaufp}{\ensuremath{\theta_{0\rightarrow\textrm{p}}}\xspace}
\newcommand{\muzpSch}{\ensuremath{\mu_{\textrm{CPD}}}\xspace}
\newcommand{\thzpSch}{\ensuremath{\theta_{\textrm{CPD}}}\xspace}

\newcommand{\fzent}{\ensuremath{{f_{\textrm{cent}}}}\xspace}    %

\newcommand{\mukrit}{\ensuremath{\mu_{\textrm{crit}}}\xspace}
\newcommand{\munkrit}{\ensuremath{\mu_{0,\,\textrm{crit}}}\xspace}

\newcommand{\SigE}{\ensuremath{\textrm{g}\,\textrm{cm}^{-2}}\xspace}

\newcommand{\kapEG}{\ensuremath{\textrm{cm}^2\,\textrm{g}^{-1}_{\textrm{gas}}}\xspace}  %
\def\kms{\ensuremath{\textrm{km}\,\textrm{s}^{-1}}\xspace}    %
\def\MPktEE{\ensuremath{\ME\,\textrm{yr}^{-1}}\xspace}        %
\def\MPktEJ{\ensuremath{\MJ\,\textrm{yr}^{-1}}\xspace}        %
\def\FEcgs{\ensuremath{\textrm{erg}\,\textrm{s}^{-1}\,\textrm{cm}^{-2}}\xspace}

\def\twb{TWA~27B\xspace}

\renewcommand{\arraystretch}{1.4}

\usepackage{booktabs}  %

\begin{document}

\title{Semianalytical Accretion-Tracer Emission: Forming Planets Are Intrinsically Faint}

\author[0000-0002-2919-7500]{Gabriel-Dominique Marleau}
\affiliation{%
Division of Space Research \&\ Planetary Sciences,
Physics Institute, University of Bern,
Gesellschaftsstr.~6,
3012 Bern, Switzerland%
}
\affiliation{%
Max-Planck-Institut f\"ur Astronomie,
K\"onigstuhl 17,
69117 Heidelberg, Germany
}
\affiliation{
Fakult\"at f\"ur Physik,
Universit\"at Duisburg--Essen,
Lotharstra\ss{}e 1,
47057 Duisburg, Germany%
}

\email{gabriel.marleau@uni-due.de}

\begin{abstract}
Direct-imaging surveys have looked for accreting planets through their accretion tracers such as \Ha but have been less fruitful than expected.
However, up to now, hydrogen-line emission at accreting planets has been estimated primarily with extrapolations of stellar-scaling relationships or with theoretical spherically-symmetric computations.
To predict the line emission intensity during the formation phase,
we follow the consequences of angular momentum conservation of the material accreting onto a gas giant in a protoplanetary disc. We focus on the limiting case that magnetospheric accretion does not occur, which yields a conservative estimate of the line emission and should correspond to certain epochs during formation.
We extend but simplify an existing analytical description of the multidimensional gas flow onto an accreting gas giant, the ballistic infall model, and combine this with detailed shock emission calculations.
Applying this to data from a global planet formation model,
we confirm that the line-emitting accretion rate is a minuscule fraction of the gas inflow into the Hill sphere. Also,
forming planets are mostly fainter than \PDSbc{} or WISPIT\,2\,b,
with a maximum \Ha line luminosity $\LLinie\sim10^{-7}~\LSonne$ roughly
independent of planet mass. Most surveys have not been sensitive to such faint planets. Other hydrogen lines in the near-IR (NIR) are fainter by 1--2~dex.
This implies that accreting planets are fainter than
from past estimates,
such that the non-detections are not as constraining as thought.
A deeper look closer in to the host stars could well reveal many forming super-Jupiters.
\end{abstract}


\section{Introduction}
 \label{Th:intro}

The discovery of accretion-tracing \Ha emission at the planetary-mass companions \PDSbc{} found in a disc around a 5-Myr star \citep{mueller18,Haffert+2019} encouraged the search for more accreting companions to young stars.
Surprisingly, these companions have remained elusive, and it is only very recently that promising additions (2M1612~b and WISPIT\,2\,b; \citealp{li25,vancapelleveen25a,close25b}) have been made to the list.
However, estimates of line luminosities,
whether meant as predictions (e.g., \citealp{mordasini17,close20}), made to analyse survey results (e.g., \citealp{Cugno+2019,Zurlo+2020,xie20,hu22,follette23,plunkett25}), or used to translate line fluxes into physical properties of individual objects, have often relied on uncalibrated extrapolations of empirical relationships between the accretion luminosity \LAkk and the line-integrated luminosity \LLinie for Classical T~Tauri Stars (CTTSs; \citealp{rigliaco12,alcal17,Komarova+Fischer2020,rogers24,fiorellino25})\footnote{While the relationships of \citet{rigliaco12} are still often used,
this paper was the
very first on a series about %
the accretion properties
at %
very low mass stars and brown dwarfs in nearby star-forming regions
with X-Shooter. The relations were refined in the next papers by \citet{alcal14,alcal17}, and recently in \citet{fiorellino25}. They have all better calibrations with more data points, and therefore supersede the work of \citet{rigliaco12}, which should not be used anymore (C.\ Manara 2023, 2025, priv.\ comm.).}.
It is usually acknowledged that this approach yields only estimates, and recent work suggests that the extrapolations might underestimate by more than an order of magnitude the accretion rate needed to generate an observed line luminosity \citep{AMIM21L,betti22b,betti22c}. Therefore, it would be useful to develop dedicated-modelling estimates that, hopefully, could be more accurate within their set of assumptions.

After the gas has made it from the large scales---be it the molecular core of a low-mass brown dwarf or the natal protoplanetary disc (PPD) of a super-Jupiter---down to the sphere of influence of a gas giant,
it is not yet entirely clear how it travels ``the last mile'' to be incorporated into the low-mass accretor.
There is possibly a difference between (effectively) isolated objects, and companions in (a gap in) the gas disc around their primary. Isolated low-mass objects almost certainly accrete only from a local reservoir, a circumplanetary disc (CPD). This suggests strongly that, if they emit accretion lines, these tracers must be coming from gas hitting the surface of the object ballistically thanks to magnetospheric accretion \citep[e.g.,][]{lovelace11,hartmann16}, with interesting implications for the magnetic field of young planets (e.g., \citealp{christensen09,hasegawa21}; \citealp{batygin25}).  %
Otherwise, the gas will be processed by a boundary layer between the CPD and the object (e.g., \citealt{kley89a,popham93}; \citealt{owenmenou16}\footnote{In that work, the estimate to justify the relevance of boundary-layer accretion inputs into the \citet{christensen09} scaling a much lower value for the luminosity of Jupiter than usually considered during formation, which leads to a different conclusion than in the by now classical computation repeated by many authors. Nevertheless, other effects could be responsible for impeding magnetospheric accretion.}), which might not lead to line emission. %
Planetary-mass objects which are companions also can be accreting by magnetospheric accretion \citep{demars23,aoyama24twa}, from which exquisite constraints on formation-phase properties can be derived at least in the special case of Jupiter \citep{batygin25}.
Companions in protoplanetary discs have further possible sources of line emission, namely from the material coming from the Hill sphere directly as it shocks on the surface of the planet or on the CPD close to the planet \citep{aoyama18,m22Schock}.

Here, we present %
estimates of the accretion-line luminosity of gas giants able to draw gas from Hill-sphere scales. We do not include explicitly magnetospheric accretion, which we justify in Section~\ref{Th:keineMagAkk}.
The set-up and line emission calculation are comparable to the one of \citet{takasao21} and \citet[][hereafter \citetalias{m22Schock}]{m22Schock}, except that we do not use the direct output of multidimensional simulations but rather develop a semianalytical approximation.
It extends and modifies a small subset of the derivations in the work of \citet{ab22} and \citet{taylor24}. %
The semianalytical approach lets us cover %
a large range of planet masses or accretion rates as relevant for planet formation \citep[e.g.][]{adams21,mordasini24} at essentially no cost.
We apply this to a whole population of forming planets, which also has the advantage of taking into account the likely correlations between the input parameters. %
We study not only \Ha but also other strong hydrogen lines.
Section~\ref{Th:setup} describes the accretion flow in the Hill sphere as well as the choice of input parameters.
Section~\ref{Th:res} presents the luminosities as a function of mass and as histograms.
In Section~\ref{Th:Scheibe} we discuss briefly some aspects beyond our model but also implications of our results before applying it to survey results, and in Section~\ref{Th:summconc} we summarise and conclude.
Appendix~\ref{Th:exakt} presents an exact solution of the flow
and Appendix~\ref{Th:fzent1/3} shows the effect of varying the centrifugal radius.

\section{Theoretical set-up}
 \label{Th:setup}

We use a set-up similar to \citetalias{m22Schock}, where we followed in 2.5D\footnote{That is, 2D polar-plane $(r,\theta)$ simulations following quantities averaged azimuthally around the planet. This can also be described as 3D simulations with only one azimuthal cell.} the accretion flow itself, emphasising the importance of not smoothing the gravitational potential when one is interested in the gas flow and thermodynamics close to the surface of the planet.
We consider a planet surrounded by a CPD and accreting from the Hill sphere.
There is a free-fall region, the planet surface with a certain radius, and a CPD with a height and extent out to \Rzent.
As seems to be common in the literature, we equate
the height of the CPD to its pressure scale height \HPzpSch, while in reality the height of the CPD, where the shock occurs, should be several times \HPzpSch. However, both are poorly-constrained input parameters, making this imprecision non-problematic in practice.
The CPD extends from the midplane ($\mu=0$, where $\mu\equiv\cos\theta$ for polar angle $\theta$) up to $\muzpSch=\cos(\pi/2-\tan^{-1}\left[\HPzpSch/\RP\right])$, where \HPzpSch is the height of the CPD at the radius where its midplane connects with the planet of radius \RP. We assume that the CPD is not flared over the line-emitting region; a different \muzpSch further away from the planet will not influence the line emission and therefore does not have to be specified in our approach.
Our goal is not to calculate the formation of a planet but rather the line emission for a given set of instantaneous values for its mass, its radius, the accretion rate onto it, and so on. These will vary over time, possibly in a complex or non-monotonic way (e.g., because of episodic accretion; \citealp{brittain20}). However, it is easy to verify that the timescale needed for the gas to fall from the PPD to the planet, and the timescale needed for the gas to come into thermal equilibrium are much shorter than the formation timescales over which the relevant planet and CPD properties change \citep{ab22}. Therefore, for line-emission calculations, we only need to choose, as an input parameter, the height of the CPD, but neither the other details of its structure nor its evolution need to be computed. We are ignoring second-order effects such as the irradiation of the incoming gas by the CPD, consistent with our assumption of ballistic infall, undeflected by thermal effects (see below).%

We present in Section~\ref{Th:accflow} the structure of the free-falling accretion flow, which yields in particular the velocity and density immediately above (before) the accretion shocks on the planet surface and CPD surface.
In Section~\ref{Th:RBdusse}, we elaborate on the boundary conditions at the Hill sphere.
Section~\ref{Th:MPkt} presents two partial accretion rates of particular interest,
and Section~\ref{Th:inpar} discusses the input parameters, including the dependence of the inflow rate on the planet and PPD properties.
Finally, in Section~\ref{Th:linem} we describe how we compute the line emission.

\subsection{Accretion flow structure}
 \label{Th:accflow}
 
Our work is inspired by a part of the work\footnote{Their overarching goal is to develop an analytical theory of planet formation. We use here, and expand upon in a different direction, only a part of the elements they used, namely the infall dynamics.} of \citet[][hereafter \ab]{ab22}, which is the first work in an series of comprehensive  %
seminanalytical derivations of the properties of growing planets and their CPDs and the associated spectra (\citealp{taylor24,taylor25,ab25}; the last hereafter \abb). That series of work starts with the widely-used \citet{ulrich76} and \citet{cassen81} model (UCM), which provide analytical infall-collapse solutions in the star-formation context.
\citet{mendoza09} generalised this elegantly %
to a finite starting radius and non-zero energy for the orbits.
The UCM model has been extended in the stellar context by \citet{shariff22} and \citet{terebey25}, for example (see references in the latter work).
We will use the results of the 2.5D radiation-hydrodynamical simulations of
\citetalias{m22Schock}
to guide our choice of boundary conditions.
Since these simulations are not global and are of limited dimensionality,
it will be sufficient for us to use, instead of the results of \citet{mendoza09} or \abb, the simpler solution of \citet{ulrich76} as used in \ab{} (see Section~2.2 of \abb).

The flow begins at a radius $r_0$, denoted by \rmax in \citetalias{m22Schock}.
In \citetalias{m22Schock} we found that the streamlines that generate emission lines all start relatively close to the pole ($\theta=0$). Therefore, we take the expressions from \ab that describe the flow and simplify them for the small-$\theta_0$ limit ($\theta_0\lesssim30\degr$) by Taylor-expanding them to second order in $\theta_0$: $\sin\theta_0\approx\theta_0$ and $\cos\theta_0\approx1-\frac{1}{2}\theta_0^2$.
This means that only points $(r,\theta)$ linked by streamlines starting at small $\theta_0$ can be used in the following approximate expressions. However, already at dozens of planetary radii, this covers all angles especially in the free-fall region since their starting $\theta_0$'s are small.

We begin with the general implicit orbit equation \AB:
\begin{equation}
 \label{Gl:orb}
  1-\frac{\mu}{\mu_0} = \zeta\left(1-\mu_0^2\right),
\end{equation}
where $\mu_{0}\equiv\cos\theta_{0}$  %
 and $\zeta\equiv\Rzent/r$,
with \Rzent the centrifugal radius, discussed in Section~\ref{Th:fzent}.
To second order in $\theta_0$ but without restrictions on $\zeta$, Equation~(\ref{Gl:orb}) can be written in the mutually equivalent forms
\begin{subequations}
 \label{Gl:muapprox}
\begin{align}
  \mu(r,\theta)     &\approx 1 - \left(\zeta+\frac{1}{2}\right)\theta_0^2,\\
  \mu_0(r,\theta) &\approx \frac{\zeta+\mu/2}{\zeta+1/2},\\
  \mu_0^2(r,\theta) &\approx \frac{\zeta-1/2+\mu}{\zeta+1/2}.
\end{align}
\end{subequations}
This
gives the starting location (at $r=r_0$) of a streamline passing through the point $(r,\theta)$.
Inserting Equation~(\ref{Gl:muapprox}) into the expressions of \ab for the velocity components (reproduced in our Equation~(\ref{Gl:vKompexakt})), %
we obtain at locations for which the starting streamline $\theta_0\ll1$:
\begin{subequations}
\label{Gl:vKomponenteapprox}
\begin{align}
  v_r      &\approx -\vFfinfty \sqrt{\frac{1+\mu+1/\zeta}{2+1/\zeta}}, \label{Gl:vrapp}\\
  v_\theta &\approx \vFfinfty \sqrt{\frac{1-\mu}{1+\mu}\frac{1}{2+1/\zeta}\left(1+\mu-\frac{1}{\zeta+1/2}\right)} \label{Gl:vthapp}\\
  v_\phi   &\approx \vFfinfty \sqrt{\frac{1-\mu}{1+\mu}\frac{\zeta/2}{\left(\zeta+1/2\right)^2}},\label{Gl:vphiapp}\\
  \vFfinfty &\equiv\sqrt{\frac{2G\MP}{r} },  %
\end{align}
\end{subequations}
where \vFfinfty is the freefall velocity from infinity.
For large $\zeta$ (close to the planet), the velocities tend to:
\begin{subequations}
\label{Gl:vGROSSzeta}
\begin{align}
v_r      & \rightarrow \vFfinfty(r)\sqrt{\frac{1+\mu}{2}}\\
v_\theta & \rightarrow \vFfinfty(r)\sqrt{\frac{1-\mu}{2}}\label{Gl:vthappGROSSzeta}\\
v_\phi   & \rightarrow \sqrt{\frac{1-\mu}{1+\mu}}\sqrt{\frac{G\MP}{\Rzent}},\label{Gl:vphiappGROSSzeta} %
\end{align}
\end{subequations}
where in Equation~(\ref{Gl:vphiappGROSSzeta}) we effectively kept half a power of $\zeta$ in $v_\phi$ but its limit corresponding to the other terms is $v_\phi\rightarrow0$, so that %
$v_r^2+v_\theta^2+v_\phi^2\rightarrow\vFfinfty^2$, as it should for energy conservation. The first square root in Equation~(\ref{Gl:vphiappGROSSzeta}) could also be written as $\tan(\theta/2)$.

Across the shock, which by definition is perpendicular to the planet or CPD surface, the azimuthal velocity component of the gas remains constant and the gas is accelerated or slowed down in the postshock cooling zone to match the $v_\phi$ of the material present. This contributes in setting how much energy is liberated (see e.g.\ Equation~(50) in \ab) but is not a shock and does not lead to line emission.
Therefore, we report $v_\phi$ only for completeness but, as in \citet{takasao21}, will not use it.

\begin{figure*}[t]
 \centering
  \includegraphics[height=0.3\textheight]{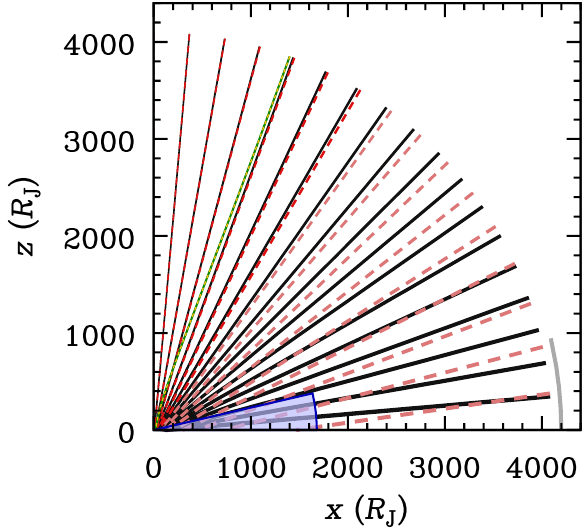}~~~~~
  \includegraphics[height=0.3\textheight]{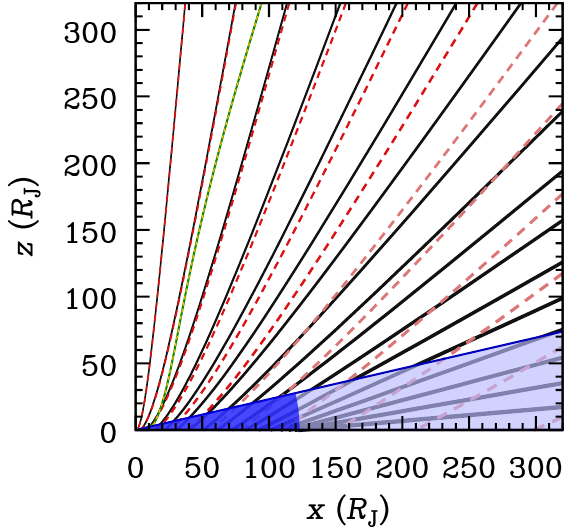}~~~~
\caption{%
Flow pattern in the poloidal plane (``side view'') for the fiducial parameters in \citetalias{m22Schock}. We set $\fzent=0.03$ (see Figure~\ref{Abb:flowducialvarfzent} for $\fzent=1/3$).
For illustrative purposes, the CPD (blue region) ends at $r=\xi\RHill$ with $\xi=0.4$ \citep{martin23}, with the region inside $\Rzent=\fzent\RHill$ highlighted (\textit{dark blue}). %
We plot implicitly the exact solution of the orbit equation (\textit{solid lines}) and
the small-$\theta_0$ solution (\textit{dashed red lines; pale}
for $\theta_0>30\degr$).  %
The \textit{green-yellow} line is the $\theta_0=20\degr$  %
streamline from \citetalias{m22Schock},
which is close to the $\theta_0=15\degr$ streamline at our $r_0=\RHill$.
Streamlines starting closer to the midplane are shown thicker to remind the reader of the density stratification (Equation~\ref{Gl:rhor0}). %
The \textit{grey arc} near \RHill displays the height of the CPD ($90\degr-\thzpSch$) and we will assume that there is in fact no inflow
over those $\theta_0$
(Sections~\ref{Th:RBdusse} and~\ref{Th:MPkt}).
}
\label{Abb:flowducial}
\end{figure*}

The orbit equation is a cubic equation in $\mu_0$ and admits an exact solution of modest complexity, which we give in Appendix~\ref{Th:exakt}. It is however easier and more insightful to work with the small-$\theta_0$ approximation. We plot in Figure~\ref{Abb:flowducial} the approximate and exact flow lines for one set of parameters defined later. The match is excellent close in to the planet and even out to several tens of Jupiter radii, %
much further out than needed to calculate shock emission \citepalias{m22Schock}.

\subsection{Boundary conditions at the outer edge}
 \label{Th:RBdusse}

Following \citet{ulrich76} and \citet{mendoza09}, \ab assumed that the density at the outer edge is constant with height. This assumption, which comes from the star-formation context, will not always hold for the super-Jupiters
that are our main object of interest, as we address after Equation~(\ref{Gl:rhor0}) below. Nevertheless, \ab discussed the relation between \RHill and \HPPPD (around their Equation~(4)). Then, \citet{taylor24} then generalised the approach of \ab by considering a non-constant vertical density profile.

The literature offers a range of predictions for the 
angular dependence of the mass influx at the outer boundary \citep[e.g.,][]{ab09b,tanigawa12,szul16,batygin18,schulik20,szul22,li23,sagynbayeva25}.  %
Consequently, \citet{taylor24} considered five different influx functions $f_i(\theta)$ which span the range of proposed geometries, from infall more concentrated at the poles to more concentrated at the equator (see also summarising schema in Figure~1 of \abb). The $f_i(\theta)$ are normalised so that the integral yield the correct total net accretion rate into the Hill sphere, which they write as \MPkthineinABT (see also Section~\ref{Th:MPkt} below): $\oint\!f_i r^2\rho v_r\,{\rm d}\Omega=\MPkthineinABT$,  that is, $\oint\!f_i\,{\rm d}\Omega=1$. Impressively, all their derivations of different quantities, also in the follow-up papers of the series \citep{taylor25,ab25}, accommodate the different $f_i$ functions\footnote{\textit{Caveat lector}: in \citet{taylor25},  %
 fortunately easily-identifiable typos have crept up into the reminder of the definitions of $f_i$.}.

At this point, it useful to estimate the size of the Hill sphere \RHill relative to the pressure scale height in the PPD $\HPPPD=a h$ at the position of the planet $a$, where $h$ is the aspect ratio of the PPD.
From the definitions of \RH and \HPPPD (e.g., \citealp{fung19}), we have
\begin{subequations}
\begin{align}
 \HPPPD &= \frac{\cs a^{3/2}}{\sqrt{G\MStern}} \\
   &= 2222  %
    \left(\frac{a}{10~\mathrm{au}}\right)^{3/2} \frac{\cs}{1~\kms} %
     \left(\frac{\MStern}{1~\MSonne}\right)^{-1/2}~\RJ\\
 \RHill &= a\left(\frac{\MP}{3\MStern}\right)^{1/3} \\
   &= 2442\,
    \frac{a}{10~\mathrm{au}}  %
     \left(\frac{\MP}{5~\MJ}\right)^{1/3} \left(\frac{\MStern}{1~\MSonne}\right)^{-1/3} 
\end{align}
\end{subequations}
and thus
\begin{subequations}
 \label{Gl:RHduerHPPPPPPcs}
\begin{align}
 \frac{\RHill}{\HPPPD} =&\, \frac{\sqrt{G}\MStern^{1/6}\MP^{1/3}}{3^{1/3}a^{1/2}\cs}\\
     =&\, 1.1 \left(\frac{\cs}{1~\kms}\right)^{-1} \left(\frac{a}{10~\mathrm{au}}\right)^{-1/2}\notag\\
     &\,\times \left(\frac{\MP}{5~\MJ}\right)^{1/3} \left(\frac{\MStern}{1~\MSonne}\right)^{1/6},
\end{align}
\end{subequations}
where $a$ is the star--planet separation and \cs is the sound speed in the midplane, which depends on the temperature profile of the PPD; we used as reference a typical value.  %
To make this estimate slightly more general, we use the prescription of \citet{chambers09} for the midplane temperature.
Where the irradiation by the central star dominates over the viscous heating in the midplane, the temperature is given by
\begin{subequations}
 \label{Gl:TirrCh09}
\begin{align}
 T &=\left(\frac{4}{7}\right)^{1/4}\left(\frac{\kB\LStern^2}{(4\pi\sigSB)^2G\mu\mH\MStern}\right)^{1/7}\frac{1}{r^{3/7}}\\
   &= 51~\left(\frac{\LStern}{1~\LSonne}\right)^{2/7} \left(\frac{\MStern}{1~\MSonne}\right)^{-1/7} \left(\frac{a}{10~\mathrm{au}}\right)^{-3/7}~\textrm{K},
\end{align}
\end{subequations}
where \MStern and \LStern are the luminosity of the star and we assume a mean molecular weight $\mu=2.353$.
At the epoch of planet formation, that is until around roughly 3--10~Myr, stars up to $\MStern\approx3~\MSonne$ are in the pre-main sequence phase \citep[e.g.,][]{dotter16}, in which the luminosity is of order $\LStern\sim0.1$--$10~\MSonne$ \citep[e.g.,][]{kunitomo17}, depending on the accretion history, and at a given age scales approximately as $\LStern\propto\MStern^p$ with $p\approx1.5$--2 (\citealp{baraffe15,dotter16,nguyen22,adams25}; the latter were reporting on the simulations of \citealt{paxton11}) for $\MStern\sim\MSonne$.
However, for flexibility, we kept \MStern and \LStern separate in Equation~(\ref{Gl:TirrCh09}).  %
Combining Equations~(\ref{Gl:RHduerHPPPPPPcs}) and~(\ref{Gl:TirrCh09}) yields
\begin{subequations}
 \label{Gl:RHduerHPPPPPPa}
\begin{align}
 \frac{\RHill}{\HPPPD} =&\, \frac{7^{1/8}4^{1/56}}{3^{1/3}}
 \left(\frac{G\mu\mH}{\kB(\pi\sigSB)^{-1/4}}\right)^{4/7}  %
 \frac{\MStern^{5/21}\MP^{1/3}}{a^{2/7}\LStern^{1/7}} \\
  =&\, 
  2.4
   \left(\frac{\LStern}{1~\LSonne}\right)^{-1/7}
   \left(\frac{a}{10~\mathrm{au}}\right)^{-2/7}\notag\\
   &\,\times\left(\frac{\MStern}{1~\MSonne}\right)^{5/21}
   \left(\frac{\MP}{5~\MJ}\right)^{1/3}.
\end{align}
\end{subequations}
This lengthscale ratio depends only weakly on the stellar parameters and, since  $1/3\approx2/7$, scales approximately as $\RH/\HPPPD\sim(\MP/a)^{0.3}$.
Thus, the ratio is larger than unity already for $\MP\gtrsim0.4~(a/10~\mathrm{au})^{6/7}~\MJ$, highlighting the importance
of considering the density stratification of the protoplanetary disc due to the stellar potential. This is most important for planets that are massive or close in\footnote{In-situ formation of (some) gas giants has been suggested as a formation channel \citep{boden00,batygin16,hasegawa19} but it remains an unsettled question \citep{ikoma25}.}. The result would have been qualitatively similar if, instead of Equation~(\ref{Gl:TirrCh09}), we had used the assumption ``minimum-mass solar nebula'' by \citet{hayashi81}, in which the PPD is transmissive (``optically thin'') radially, so that the gas temperature is given by the equilibrium with the unattenuated stellar irradiation, $L=4\pi a^4 \aST c T^4$,
where $\aST = 4\sigSB/c$ is the radiation constant. This leads to
\begin{equation}
 T_{\mathrm{H}81} = 88.1\,\left(\frac{\LStern}{1~\LSonne}\right)^{1/4} \left(\frac{a}{10~\mathrm{au}}\right)^{-1/2}~\textrm{K},
\end{equation}
which is considered part of the ``minimum-mass solar nebula''.
At tens of astronomical units, this likely overestimates the midplane temperature somewhat, as comparison with Equation~(\ref{Gl:TirrCh09}) suggests (see Figure~3 of \citealt{emsen21a} for an example).%

Therefore, as in \citetalias{m22Schock}, we take the finite $\RHill/\HPPPD$ into account by setting the density at the outer edge $r_0 = \rmax=\RHill$ to
 \begin{equation}
\label{Gl:rhor0}
  \rho_0(\theta_0) \equiv\rho(r_0,\theta_0)  %
  = \rhoMitt\exp\left[-\frac{1}{2}\left(\frac{r_0\cos\theta_0}{\HPPPD}\right)^2\right]    %
 \end{equation}
with \rhoMitt the midplane density in the PPD.
Using the fiducial values in Equation~(\ref{Gl:RHduerHPPPPPPa}), the density at the pole relative to the midplane is $f_\rho = \rho_0(0)/\rhoMitt\approx 0.06$, or a suppression by a factor of~20.
Because of the exponential, the value will depend on the details but this suggests that for massive planetary-mass accretors forming at some distance from the star, the vertical structure of the PPD needs to be considered.

\begin{figure}[t]
 \centering
  \includegraphics[width=0.47\textwidth]{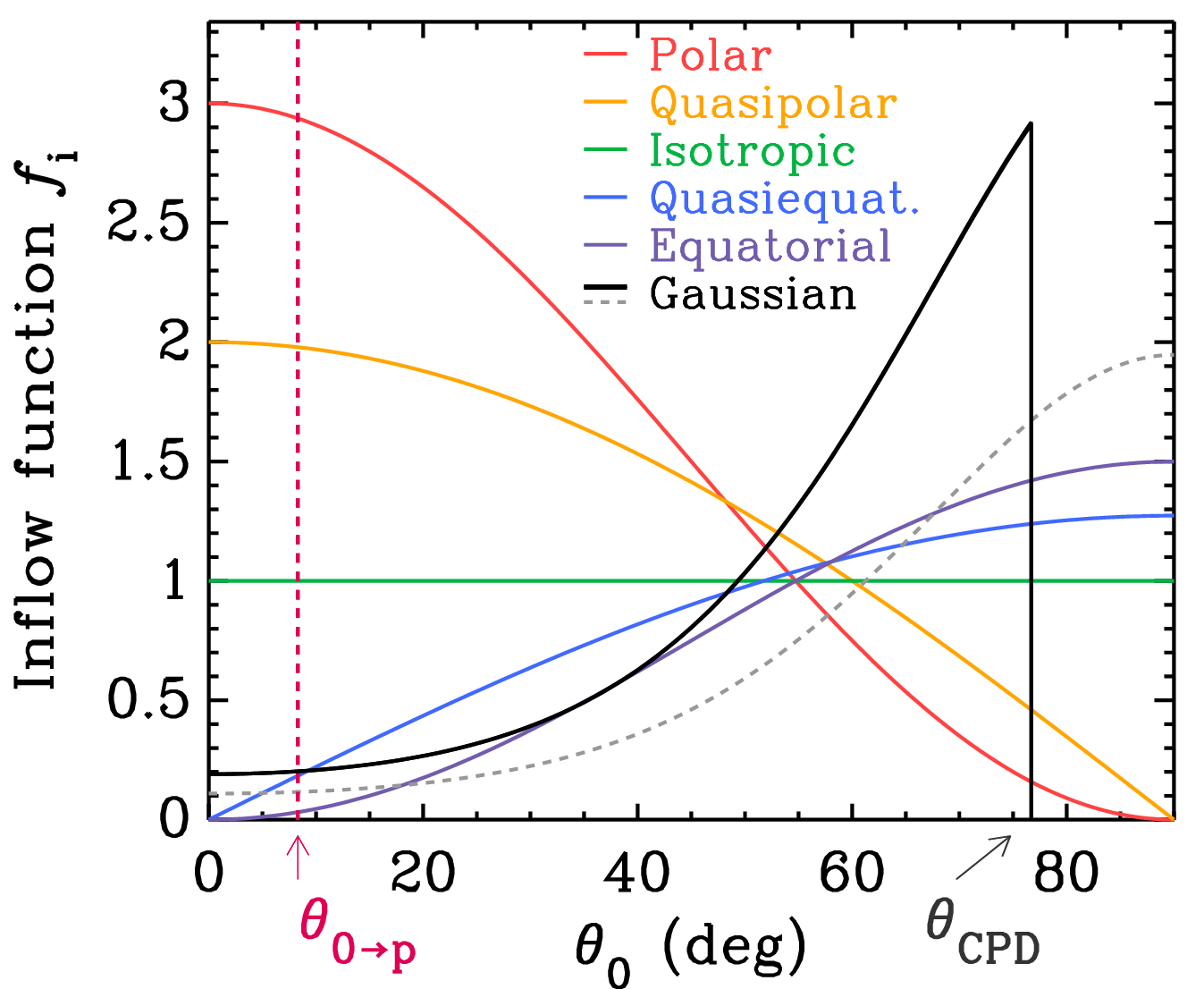}
\caption{%
Inflow functions (that is, mass loading at the Hill sphere) as in \citet{taylor24} (\textit{coloured lines}) and with an example of our adaptation for massive accreting planets (``Gaussian''; $\RHill/\HPPPD=2.4$; see Equation~(\ref{Gl:RHduerHPPPPPPa})),
either truncating the inflow approximately at the thickness of the CPD (\textit{black line}) or allowing it down to the midplane (\textit{grey dashed}).
Only the gas between the pole ($\theta_0=0$) and $\thnaufp=8.3\degr$ (Equation~(\ref{Gl:mu0min}), for the reference values in Equation~(\ref{Gl:RHduerHPPPPPPa}) and $\RP=2~\RJ$) hits the planet surface directly, with the rest landing on the CPD.
}
\label{Abb:fi}
\end{figure}

In Figure~\ref{Abb:fi}, we compare to the inflow functions of \citet{taylor24} the one resulting from taking the density stratification in the PPD into account. At the outer edge, the velocity is nearly purely radial (as Figure~\ref{Abb:flowducial} shows; $r_0\gg\Rzent$), so that $v_0(\theta_0)$ is essentially constant and  $f_i\propto\rho_0$ only.
We also truncate the mass inflow at a finite angle above the midplane, at a polar angle $\thzpSch\equiv\arccos(\muzpSch)$, as we detail below in Section~\ref{Th:MPkt} (represented schematically in Figure~\ref{Abb:flowducial}).
Already the  %
non-truncated inflow function but even more so the nominal one show much more inflow close to the equator than even the ``Equatorial'' scenario in \citet{taylor24}. This will have interesting consequences which we discuss later.

Equation~(\ref{Gl:rhor0}) is approximate because it assumes axisymmetry and hydrostatic equilibrium at the Hill radius but also because it does not include the changes to the pressure and temperature structure due to the planet (e.g., through accretion heating).
However, we let the midplane density \rhoMitt{} at the position of the planet,
\begin{equation}
 \rhoMitt = \frac{\Sigma}{\sqrt{2\pi}\HPPPD},
\end{equation}
reflect the gap opening by the planet in that we set $\Sigma$ to the reduced surface density in the gap, for instance according to \citet{kanagawa17}.
This $\Sigma$ can be related to an unperturbed surface density $\Sigma_0$ by choosing a viscosity $\alpha$, or one can take $\Sigma$ itself as the input quantity instead of $(\Sigma_0,\alpha)$ and a model for the surface-density reduction in the gap.

To calculate line emission from the shock requires the preshock densities. Since the density is a known (chosen) function at the outer edge, it is easiest to use mass conservation along a mass ``tube'' from the outer edge down to the planet and the CPD to obtain the densities there.
Recalling the notation $\rho_0(\theta_0)\equiv\rho(r_0,\theta_0)$ for the density at the outer edge, we have in general
\begin{equation}
 \label{Gl:rhogen}
  \rho(r,\theta) = \rho_0(\theta_0) \frac{r_0^2 |v_r(r_0,\theta_0)|}{r^2|v_r(r,\theta)|} \frac{d\mu_0}{d\mu},
\end{equation}
where $\theta_0=\theta_0(r,\theta)$.
Taking the implicit derivative of the general orbit equation (Equation~(\ref{Gl:orb})) and then using the approximate orbit equation, one can rewrite the last factor in Equation~(\ref{Gl:rhogen}) as
\begin{subequations}
\begin{align}
  \frac{d\mu_0}{d\mu} &= \left(\left[3\mu_0^2-1\right]\zeta+1\right)^{-1}\label{Gl:dmu0dmu}\\
  & \approx\frac{1+1/(2\zeta)}{3\mu-1+2\zeta+1/(2\zeta)}\label{Gl:dmu0dmuapp}\\
  &\rightarrow \frac{r}{2\Rzent}, \label{Gl:dmu0dmuapp^2}
\end{align}
\end{subequations}
where Equation~(\ref{Gl:dmu0dmu}) is exact and is the result quoted in \ab{} (see our Figure~\ref{Abb:thdm0}). Equation~(\ref{Gl:dmu0dmuapp}), which indeed correctly contains $\mu$ and not $\mu_0$, holds to second order in $\theta_0$ for $\theta_0\ll1$ but still for arbitrary $\zeta$.
The limit in Equation~(\ref{Gl:dmu0dmuapp^2}) assumes that $\zeta$ is large
(which in turn implies that $\theta_0$ is small).
In the latter limit, for instance close to the planet, Equation~(\ref{Gl:dmu0dmuapp^2}) makes it clear that a small range in starting angles is responsible for accretion onto a larger range of angles: $d\mu_0\ll d \mu$ if $\Rzent\gg\RP$.

\subsection{Accretion rates}
 \label{Th:MPkt}

A useful diagnostic quantity is \MPktPdir, the total mass flux hitting the planet directly.
The interest of \MPktPdir is twofold: it is the minimum, ``guaranteed'' mass growth rate of the planet, as opposed to the mass that falls onto the CPD and that might move away from the planet. Secondly, \MPktPdir estimates within a factor of a few the mass infall rate that generates emission lines \citepalias{m22Schock}.

One can calculate \MPktPdir by integrating at \RP the local mass flux
\begin{equation}
 \label{Gl:mPunkt}
  \mPunkt=4\pi r^2\rho |v_r|
\end{equation}
over the free surface of the planet or at $r_0$ over the corresponding streamlines. The latter is easier, and we write:
\begin{equation}
 \label{Gl:MPktPdirexakt}
  \MPktPdir = \int_{\munaufp}^1 4\pi r_0^2\rho_0(\theta_0)|v_r(r_0,\theta_0)|\,d\mu_0.
\end{equation}
The lower limit of the integral $\mu_0=\munaufp$ is connected to \muzpSch at \RP and can be obtained from the Taylor-expanded expression for the streamlines (Equation~(\ref{Gl:muapprox})):
\begin{subequations}
\label{Gl:mu0min}
\begin{align}
  1-\munaufp &= \frac{1-\muzpSch}{2}\frac{\RP}{\Rzent}\\
    \thnaufp &= \sqrt{ \left(1-\muzpSch\right)\frac{\RP}{\Rzent} }.
\end{align}
\end{subequations}
The height of the ``polar cap'' at the Hill radius in Equation~(\ref{Gl:MPktPdirexakt}) is small: $\Delta z=(1-\munaufp)\RHill\ll\HPPPD$ (Equation~(\ref{Gl:RHduerHPPPPPPa})).
Therefore, the $\rho_0$ factor in Equation~(\ref{Gl:MPktPdirexakt}) is roughly constant and can be evaluated at the pole, and $v_r(r_0,\theta_0)\approx\vFfinfty(r_0)$ to an even better approximation. This yields
\begin{subequations}
\label{Gl:MPktPdirapprox}
 \begin{align}
  \MPktPdir \approx &\,4\pi\RHill^2\rho_0(0)\vFfinfty(\RHill)\left(1-\munaufp\right)\\
            \approx &\,2\, \Sigma\, \sqrt{\pi G\MP\RHill}\frac{(1-\muzpSch)}{\HPPPD}\frac{\RP}{\fzent}\notag\\
            & \;\times \exp \left[-\frac{1}{2}\left(\frac{\RHill}{\HPPPD}\right)^2\right],
 \end{align}
\end{subequations}
where $\fzent\equiv\Rzent/\RHill$ is an important parameter discussed below.
Equation~(\ref{Gl:MPktPdirapprox}) is an excellent approximation, as we verified numerically, and it shows that for fixed other parameters, $\MPktPdir\propto\Sigma/\fzent$ or $\MPktPdir\propto\Sigma/\Rzent$.

Another noteworthy, and more fundamental, accretion rate is the net mass flux entering the Hill sphere, \MPktnettoHill. Hydrodynamical simulations predict it to be related to the surface density of the PPD a few Hill radii away from the planet \citep{tanigawa02}.
Three-dimensional simulations show the presence of a net outflow away from the planet in the disc midplane (e.g., \citealp{tanigawa12,cilibrasi23,martin23}), at least for some CPD aspect ratios. To emulate this crudely, we truncate the solution at $r_0$ between the midplane and \muzpSch, effectively setting $v=0$ there, that is, not letting any material enter the Hill sphere close to the midplane.
The reasoning is that the 
horseshoe orbits might have an aspect ratio comparable to that of the CPD,
as the simulations of \citet{machida08} suggest, for instance.
However, \muzpSch is determined for the regions closest to planet and might not be related to the azimuthally averaged dynamics between \Rzent and \RHill.
Therefore, it is not clear how exactly this truncation should be, and the current approach is meant only as an acceptable starting point.
We verified that integrating down to the midplane or only to \muzpSch changes the mass influx by a factor of 0.2--0.7 over 1.5~dex in mass, which is modest.  %

Integrating $\mPunkt$ (Equation~(\ref{Gl:mPunkt})) at \RHill, the net mass flux into the Hill sphere is
\begin{subequations}
\label{Gl:MPktnettoHill}
 \begin{align}
  \MPktnettoHill =&\, \int_{\muzpSch}^14\pi r_0^2 \rho_0(\theta_0)|v_r(r_0,\theta_0)|\,d\mu_0\\
      \approx&\, 4\pi\RHill^{3/2} \sqrt{2G\MP}
        \int_{\muzpSch}^1 \rho_0(\theta_0)
         \,d\mu_0 \label{Gl:MPktnettoHillapprox-1}\\
     \approx &\,2\pi \Sigma \sqrt{2G\MP\RHill}\times\Derf\label{Gl:MPktnettoHillapprox},
 \end{align}
\end{subequations}
where we have used that $r_0=\RHill$ and defined
\begin{subequations}
 \begin{align}
\Derf &\equiv\erf[x(1)]-\erf[x(\muzpSch)],\\
       x(\mu_0) &\equiv\frac{\mu_0}{\sqrt{2}}\frac{\RH}{\HPPPD}
          = \mu_0\sqrt{\frac{\RH^2\RB}{2a^3q}}.
 \end{align}
\end{subequations}
Equation~(\ref{Gl:MPktnettoHillapprox-1}) is approximate but Equation~(\ref{Gl:MPktnettoHillapprox}) follows exactly from it.
Since the choice of the lower integration limit to define \MPktnettoHill is somewhat arbitrary, we simplified $v_r$ to be equal to \vFfinfty for all angles. The exact solution is indeed not too far from radial (Figure~\ref{Abb:flowducial}), but in general we are effectively integrating down to a location slightly closer to the pole than \muzpSch.  %
This slight approximation made it possible to obtain a semianalytical answer involving the error function.
Since $x(1)$ is neither very small nor very large,
the factor \Derf{} does not seem simplifiable, but it will turn out to be of order unity.  %
The important property of \MPktnettoHill is that it scales linearly with $\Sigma$, which we will use extensively.

For comparison, \ab define the accretion rate of material that stays within the Hill sphere \MPkthineinABT, to which our \MPktnettoHill corresponds, as an integral over all angles, that is, down to the midplane, instead of down to the CPD height. This too is a valid assumption given the uncertainties in the inflow geometry as mentioned in Section~\ref{Th:accflow}.  %
\subsection{Input parameters}
 \label{Th:inpar}

Our model so far has several input parameters: $(\MP,\RP,a,\Sigma,h,\MStern)$,
using the same parametrisation as in \citetalias{m22Schock},  %
as well as the  %
ill-constrained parameters \fzent and \muzpSch.
We could study how the line luminosities depend on each parameter when varied separately.
However, several studies clearly predict that the ``accretion rate'' \MPkt (we will come back later to the meaning of this term) depends on the planet mass, and linearly on the surface density.
Therefore, we review some dependencies of \MPkt on \MP and use this to choose a meaningful scaling for the surface density $\Sigma$.
Then, we address \fzent.

\subsubsection{Accretion rate scalings}

There are different prescriptions in the literature that predict \MPkt from the properties of the protoplanetary disc around a forming planet.
Including our result (Equation~(\ref{Gl:MPktnettoHill})), they can be written as follows:
\begin{equation}
 \label{Gl:MPktUebersicht}
 \frac{\MPkt}{\Sigma_0\Omega \RH^2} =
 \begin{cases}
  \left(\frac{q}{3}\right)^{-2/3}\FBoden & \mbox{\citep{boden13}} \\
  \frac{1}{8}\fLuecke           \approx0.13\fLuecke & \mbox{(\citetalias{emsen21a}, ``max,\,2D'')} \\
  \frac{9}{\sqrt{2\pi}}\fLuecke \approx4\fLuecke & \mbox{\citep{choksi23}} \\
  \underbrace{2\pi\sqrt{6}\,\Derf}_{\approx\,5\mbox{--}10}\,\fLuecke & \mbox{(this work)},
\end{cases}
\end{equation}
with the short form \citetalias{boden13} for \citet{boden13}
and \citetalias{emsen21a} for \citet{emsen21a}, the first paper of the ``New Generation Planetary Population Synthesis'' series on the third-generation Bern planet formation model. We note the following:
\begin{itemize}
\item
The non-constant factor $\FBoden=\FBoden(\alpha,q,h)$ comes from Equations~(A5) and~(A6) of \citetalias{boden13} but multiplied by 0.15 to reproduce their Figure~12a (C.~Mordasini 2019, private communication).

\item 
The ``max,\,2D'' scaling of \citetalias{emsen21a} is approximate and is a maximum for superthermal planets,
for which the ``thermal mass ratio'' $\qth\equiv q/h^3=3(\RH/\HPPPD)^3>1$, where $q\equiv\MP/\MStern$ \citep[e.g.][]{korycansky96,machida08}.
In the form written in Equation~(\ref{Gl:MPktUebersicht}), the scaling holds when the relative-velocity factor is equal to $\Omega\RH$ and not \cs (see the expression in \citealt{emsen21a}, or \citealt{choksi23}).

\item
Similarly, we selected the $\qth\gtrsim10$ scaling of \citet{choksi23}.
They based the scaling on the same analytical argument (their ``Hill,\,2D'' case) but corrected it by an empirical factor from their simulations.
This large-\qth{} limit is the appropriate one for the planets in which we are interested, as mentioned above.  %

\item 
``This work'' refers to Equation~(\ref{Gl:MPktnettoHillapprox}), which is the ballistic model of \ab combined with our density structure and choice for the lower integration limit.
Typically, the term $\Derf\approx0.3$--0.6 when applied to the population synthesis planets
that we will introduce below.

\end{itemize}

The meaning of \MPkt is slightly different between the different works in Equation~(\ref{Gl:MPktUebersicht}):
in \citetalias{boden13} and NGPPS it is interpreted as \MPktnettoHill,
whereas \citet{choksi23} write that their \MPkt is only the inflow component into the Hill sphere and thus an upper limit on \MPktnettoHill.
In our work, \MPkt is the actual net rate \MPktnettoHill.
How this relates to the actual growth rate of a planet at a given time is an entirely different question that needs to be addressed separately.  %

In Equation~(\ref{Gl:MPktUebersicht}), we scaled \MPkt by the unperturbed surface density $\Sigma_0$ because the \citet{boden13} formula includes the surface density reduction due to gap opening,
whereas in the other studies (including the present one) the scaling was derived for the local, that is, reduced, $\Sigma$.
This is why, for all but \citetalias{boden13}, we included a gap reduction factor $\fLuecke=\fLuecke(\alpha,q,h,\ldots)$ that needs to be determined by a separate modelling effort.

Here, we use the results of \citetalias{emsen21a}, which yields a distribution of planet and PPD properties as a function of time.
This is meant to provide guidance as to possible correlations between the parameters when taking several physical effects into account.
We use the main, 100-embryo-per-PPD simulation ``NG76''\footnote{%
Snapshots with restricted columns are available for plotting and downloading at \url{https://dace.unige.ch/populationAnalysis/?populationId=ng76}.}.
To have some self-consistency in the geometrical picture of our model
and avoid the weakest accretors,
we keep from the snapshots only the synthetic planets that match the following criteria:
 $a>10~\RStern$,
 $\MPktPopsynth>10^{-10}~\MPktEE$,
 $\MP>0.5~\MJ$,
 $\vFfinfty>27~\kms$,
 $0.03~\RHill>5~\RP$,
where $a$ is the semimajor axis, \RStern the stellar radius, and
\MPktPopsynth the accretion rate in the population synthesis.
The last criterion ensures that the smallest centrifugal radius that we will consider, 
with $\fzent=0.03$, still leaves a minimal amount of room for a CPD.
We take the population snapshots at $t=1$, 2, 3, 4, and 5~Myr,
in which respectively $(221,207,144,93,64)$ planets match these criteria.
They all occupy relatively similar areas of parameter space.
To have better statistics, we ignore the time coordinate and study them together.
This is consistent with the typical uncertainty of a few~Myr on the age of young systems.

\begin{figure}[t]
 \centering
  \includegraphics[width=0.47\textwidth]{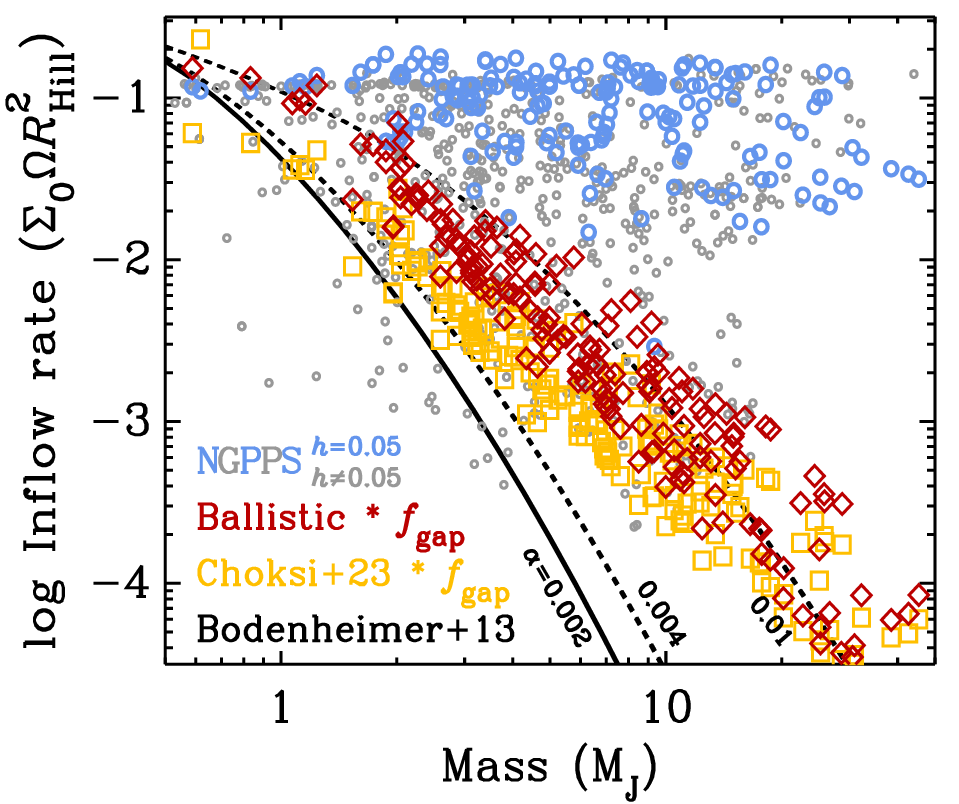}
\caption{%
Comparison of mass inflow rates (``accretion rates'') into the Hill sphere from different work, normalised to the local value of $\Sigma_0\Omega\RHill^2$:
the actual rate in NGPPS (blue circles: $h\approx0.05$; grey: $h\napprox0.05$),
the scaling from the ballistic model in this work (red diamonds), and
the \citet{choksi23} ``2D-Hill'' scaling (yellow squares).
The latter two need only the $q$ and $h$ values from the NGPPS planets
as input into Equation~(\ref{Gl:MPktUebersicht}) to determine the reduction of $\Sigma$.
The \citet{boden13} scaling, derived for $h=0.05$,
includes the effects of gap opening and is shown as black lines (solid for $\alpha=0.002$ and dashed for~0.004 and~0.01).
}
\label{Abb:MPktUebersicht}
\end{figure}

In Figure~\ref{Abb:MPktUebersicht}, we show the accretion rates predicted by the prescriptions of Equation~(\ref{Gl:MPktUebersicht}).
For \fLuecke, needed for \citet{choksi23} and our scaling, we use the formula from \citet{kanagawa17} as a popular choice but note that some alternatives, such as \citet{zhang18}, exist.
We use the population synthesis planets to have a distribution of $(q,h)$ values for these two models.
However, the \citetalias{boden13} scaling depends in an unknown way on $h$ because their simulations were done for only one value, $h=0.05$. Therefore, we highlight the synthetic planets with $h=0.04$--0.06. %
The spread in the ``ballistic'' (this work; red) and \citet{choksi23} scalings (yellow) come from the high sensitivity of $\fLuecke$ on $h$ ($\fLuecke\propto h^5$ for high $q$). Otherwise, they are similar, as Equation~(\ref{Gl:MPktUebersicht}) suggested.
The fit \FBoden of \citetalias{boden13} is shown for $\alpha=0.004$ %
but also, extrapolated linearly in $\alpha$ from their fits for $\alpha=0.004$ and $0.01$, for the value of $\alpha=0.002$  %
that was used in \citetalias{emsen21a}.

By contrast, compared to the other works, the accretion rates in \citetalias{emsen21a} are similar at $\MP\approx1~\MJ$ but higher by a factor of ten or more already at $\MP=3~\MJ$,
and of 100--300 at 10~\MJ.
The spread in scaled accretion rates from \citetalias{emsen21a}, and the reason why they are high relative to the other scalings, has two aspects.
One is that in \citetalias{emsen21a}, the surface density at the location of the planet is not reduced (that is, $\fLuecke=1$) to compute the accretion rate.
This is an optimistic limit that is based on the eccentric instability \citep{pap01,kley06}.
This would place all the points at $\MPkt/(\Sigma_0\Omega\RHill^2)=1/8$ (Equation~(\ref{Gl:MPktUebersicht})) in Figure~\ref{Abb:MPktUebersicht},
where there is a indeed a certain pile-up in the grey points, which have $h\napprox0.05$ (for most synthetic planets, $h$ is smaller).
The second aspect is that for many planets, what limits the accretion is the mass available per timestep in the feeding zone \citep[][their Section~4.1.2]{emsen21a}, which is (unfortunately) a numerical consideration (A.~Emsenhuber 2023, private communication).
Thus many planets are accreting more slowly than the expected scaled rate $\MPktPopsynth=(1/8)\Sigma_0\Omega\RHill^2$, yet still clearly higher than in the other prescriptions. The few planets that are accreting faster (however, by only at most some tens of percent) than the value of Equation~(\ref{Gl:MPktUebersicht}) are presumably not quite in the limiting regime assumed to derive Equation~(\ref{Gl:MPktUebersicht}).

The conclusion from this comparison is that at fixed $\alpha$,
the prescription of \citet{boden13} yields a scaled accretion rate $\propto\MPkt/\Sigma$ that is the the smallest,
whereas in \citetalias{emsen21a} the scaled rate is orders of magnitude higher.
The scalings from \citet{choksi23} and our model are intermediate.
We next turn to describe how we use this to set $\Sigma$ for our estimate.

\subsubsection{Surface densities}
 \label{Th:skalSigma}

As input into our model, we set $\MStern=1~\MSonne$ and take the population synthesis values of $(\MPktPopsynth\equiv\MPktvekst, \RP, a, h)$ for each synthetic planet. We then choose $\Sigma$ to have \MPktnettoHill from our ballistic model be equal to \MPktPopsynth.
This is a limiting case that assumes that all the mass entering the Hill sphere is added quickly to the planet:
\begin{equation}
 \label{Gl:MPktvekstglychHill}
\MPktvekst=\MPktnettoHill.
\end{equation}
Since the mass in the CPD is very small\footnote{Analysing the simulations of \citetalias{m22Schock}, we find
$\MzpSch\approx0.02~\ME$ or $\MzpSch=0.5~\ME$
for the main $\MP=2$ or $5~\MJ$ simulations, respectively.
The surface densities $\Sigma\sim10^2$--$10^3$~\SigE are radially relatively flat.},
this limit requires the transport time through the CPD to be short,
which should hold in the limit of sufficiently high viscosity in the CPD (\citealp{papnel05}; \ab; \citealp{lesur23ppvii}; \abb).
The other limiting scenario would be that only what falls directly onto the planet (\MPktPdir) contributes to its growth, which we do not use here.
Thus, we set as surface density
\begin{equation}
 \label{Gl:SigSkal}
 \Sigma = \frac{\MPktPopsynth}{(\MPktnettoHill/\Sigma)},
\end{equation}
where $\MPktnettoHill/\Sigma$ is independent of $\Sigma$ and is computed with the integral in Equation~(\ref{Gl:MPktnettoHill}).
We fix $\muzpSch=0.23$, which  %
 influences \MPktnettoHill only modestly.
We do not need to model explicitly gap opening, which would require choosing an $\alpha$;
as mentioned above,
the $\Sigma$ that enters our model is the actual (reduced) surface density at the position of the planet.
\begin{figure}[t]
 \centering
  \includegraphics[width=0.47\textwidth]{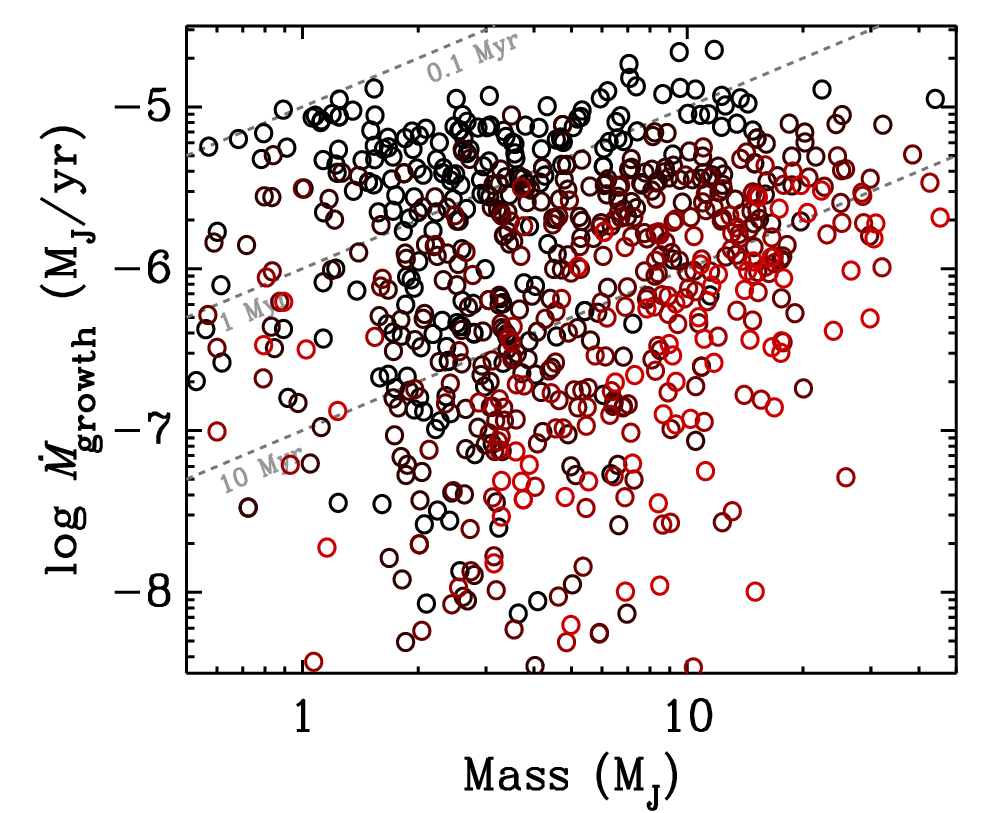}\\    %
  \includegraphics[width=0.47\textwidth]{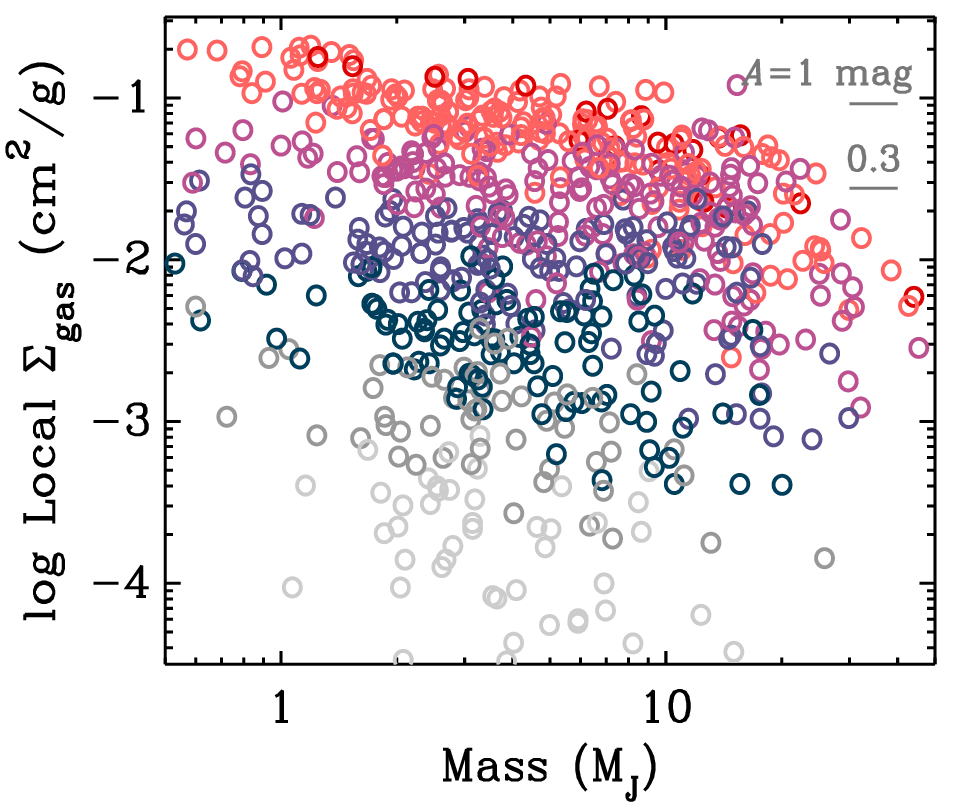}  %
\caption{%
\textit{Top}: Planet growth rates in \citetalias{emsen21a} for our selected planets at 1--5~Myr (black to red). Dashed lines show characteristic growth times of 0.1, 1, 10~Myr.
\textit{Bottom}:
Gas surface density at the location of each planet,
chosen based on the $q$ and $h$ parameters of the \citetalias{emsen21a}
so that our \MPktnettoHill %
equal the planet growth rate in the population synthesis \MPktPopsynth (Equation~(\ref{Gl:SigSkal})).
Colours indicate
$\log\MPktPopsynth/(\MPktEJ)$ in bins with edges at $\{\infty,-5,-5.5,\ldots,-7.5,-\infty\}$ (dark red, salmon, lilac, purple, dusty blue, grey, pale grey).
Short line segments on the right indicate the $\Sigma$ for which the vertical extinction $A=1$~or 0.3~mag when $\kappa=10~\kapEG$ (see text).  %
}
\label{Abb:Sigma}
\end{figure}

In Figure~\ref{Abb:Sigma}, we show the planet growth rate \MPktvekst from the population synthesis and the (reduced) surface density $\Sigma$ in the main part of the gap (Equation~(\ref{Gl:SigSkal})).
The maximum \MPktvekst values are relatively independent of planet mass.
The corresponding surface densities go up to $\Sigma\sim0.1~\SigE$, without notable features in the distribution down to $\Sigma\sim10^{-3}~\SigE$.

The extinction at any wavelength can be estimated as $A=\ln(10)/2.5\times\tau\sim\Sigma\kappaStbint\fpg$, where $\tau$ is the optical depth, \kappaStbint is the intrinsic material opacity, and \fpg is the dust-to-gas ratio. \citet{maea21} attempted to estimate this from the literature and found $\kappaStbint\fpg\sim10^2$--1~\kapEG at \Ha. Thus only for $\Sigma\gtrsim0.1~\SigE$ (indicated by a segment in Figure~\ref{Abb:Sigma}) will the extinction at \Ha be larger than 1~mag when using the upper $\kappa$ value.

Therefore, for most planets, extinction might not be important at \Ha since $\tau\lesssim0.1$. Near-infrared (NIR) hydrogen transitions are likely even less affected since dust opacity usually decreases with wavelength (e.g., \citealt{gordon23}). Gas opacity, which is very non-monotonically temperature- and wavelength-dependent, could be higher or lower at other wavelengths, but \citet{maea21} estimated that the gas accretion column was optically thin at \Ha. Their assumed cone-wise spherically symmetric geometry in fact meant that their extinction estimates are upper limits.
We will discuss that this implies that extinction is likely not the dominant effect in explaining why so few planets have been caught accreting so far, and that faint emission is likelier.  %

Studying the AS~209 system, \citet{cugno25} were able to estimate the extinction in a wide gap and found that it was worryingly both non-negligible ($A_V\approx5$~mag) and grey, that is, not strongly decreasing with wavelength, and also not spatially constant in the gap. The arguments we have just made do not contradict this because \citet{maea21} were studying only the extinction by the inflowing material, \textit{assuming} the residual material in the gap essentially to be optically thin. Also, as \citet{cugno25} mention, PDS~70 for instance has a (much) deeper gap than AS~209. Thus the Cugno effect could be a severe one for individual systems but it is reasonable to expect that it will not affect all, requiring another explanation for the lack of known line-emitting accretors.

\subsubsection{Centrifugal radius}
 \label{Th:fzent}

In our approach,  %
the centrifugal radius $\Rzent=\fzent\RHill$ is up to now unconstrained, yet it plays an important role in setting the line flux.
Classically, the centrifugal radius is defined by the angular momentum bias $\ell$, with $\Rzent=\ell^2\RHill/3$ (e.g., \citealp{ward10}; \abb). From simple considerations, the classical estimate for the centrifugal radius relative to the Hill radius is $\fzent=\Rzent/\RHill=1/3$ \citep{quillen98},
so that $\ell=1$, whereas averaging azimuthally the unperturbed Keplerian shear flow\footnote{This is what \citetalias{m22Schock} did for the outer boundary condition. Their Equation~(3) is in the rotating frame of the planet; compare to $\bar{j}_{\mathrm{rot}}$ of \citet{ward10}, below their Equation~(10).} gives $\ell=1/4$ \citep{lissauer91,ward10}.
A related interesting topic is the extent of the CPD, which might be a factor $\xi=0.4$ of \RHill \citep{martin11a}, adopted by \ab, while thick CPDs with aspect ratios above $\HPzpSch/r\approx0.3$ might not really be truncated \citep{martin23}. However, the extent of the CPD has no influence on our line-emission calculations.

Roughly,
gas within \Rzent has sufficiently little angular momentum to accrete onto the planet, whereas gas outside \Rzent spreads outward.
More precisely, the transition between in- and outflow occurs at a radius $r\equiv u_0\Rzent$ with $u_0\approx0.3$--1 (\abb), for polar- to equatorially-concentrated infall geometries, respectively (see Section~\ref{Th:RBdusse}); in our case, $u_0$ is probably of order unity.
If $\Rzent\gg\RP$, the velocities immediately before the shock do not depend on \Rzent (see Equation~(\ref{Gl:vGROSSzeta})). However, \Rzent does scale the preshock density linearly (Equation~(\ref{Gl:dmu0dmuapp^2})).

We draw on the simulations of \citetalias{m22Schock} to estimate \fzent. We first scale our $\Sigma$ to match the influx \MPktnettoHill in those simulations (Equation~(\ref{Gl:SigSkal})) and then choose \fzent to match \MPktPdir. This is straightforward since \MPktnettoHill does not depend on \fzent.
Fixing $\muzpSch=0.23$ for simplicity,
we find $(\Sigma=0.021~\SigE,\fzent=0.026)$ for the 2-\MJ simulation
and $(0.022~\SigE,\fzent=0.030)$ for the 5-\MJ simulation,
choosing $\RP=2.1~\RJ$ for both.
Those values $\fzent\approx0.03$ are one order of magnitude smaller than the canonical estimate $\fzent=1/3$.
A small value for \fzent agrees with the approximate scaling of \citet{ward10},
\begin{subequations}
\begin{align}
 \ell 
   &\approx0.12\sqrt{\frac{\RB}{\RHill}} + 0.13\\
   &\approx0.144\qth^{1/3}+0.13,
\end{align}
\end{subequations}
using $\RB^3/\RH^3=3\qth^2$ for the second line (\citealp{fung19}). The fiducial parameters of \citetalias{m22Schock} ($\qth=7.1$) yield $\ell\approx0.4$ or $\fzent=\ell^2/3\approx0.05$.
This agrees qualitatively with
\ab, who pointed out that \fzent is not well known but probably likely smaller than the canonical value, and with \abb specifying that this reduction might be around $\ell^2\approx1/9$--$1/4$.

For simplicity, and because the flux scales only linearly inversely with \fzent, for the whole population we will fix $\fzent=0.03$.  %
In Appendix~\ref{Th:fzent1/3}, we show that using a constant value $\fzent=1/3$ instead does not make a qualitative difference.

\subsection{Emission line calculation}
 \label{Th:linem}

To compute the line emission, we assume two contributions: from the free planet surface (from the pole down to $\mu=\muzpSch$) and from the CPD surface. The local microphysical emission model, applicable to both regions, is from \citet{aoyama18} and requires as inputs the preshock velocity perpendicular to the shock \vVorSch as well as the preshock density \rhoVorSch.
For the planet-surface shock, the relevant velocity is $\vVorSch=|v_r|$, and for the CPD it is $\vVorSch=v_\theta$ since the CPD surface is assumed to be radial at $\mu=\muzpSch$ (that is, not flared), which is a good approximation close to the planet \citepalias{m22Schock}. In both cases, the other input quantity, the ``number density\footnote{It has the units of a number density but is not the usual definition.} relative to the number of hydrogen protons'', is simply $\nVorSch=X\rhoVorSch/\mH$, with $X=0.25$ the hydrogen mass fraction and $\mH$ the atomic mass of hydrogen.

With \vVorSch from Equations~(\ref{Gl:vrapp}) or (\ref{Gl:vthapp}) for the planet or the CPD surface, respectively,
and \nVorSch from Equations~(\ref{Gl:vrapp}), (\ref{Gl:rhogen}), and~(\ref{Gl:dmu0dmuapp}), we compute the local line fluxes $F(\nVorSch,\vVorSch)$ for each transition of interest. We integrate and keep track separately of the luminosities from the free planet surface and from the CPD surface, summing both hemispheres in both cases:
\begin{subequations}
\begin{align}
 \LPlObfl &= 4\pi\RP^2 \int_{\muzpSch}^1 F(n(\RP,\theta),|v_r|)\,d\mu,\\
 \LzpSch  &= 4\pi\sin\muzpSch \notag\\
 &\quad\quad\,\times\int_{\RP} r F(n(r,\muzpSch),v_\theta)\,dr.  %
\end{align}
\end{subequations}
Along the CPD surface, we integrate out to %
slightly further out than where $v_\theta$ drops below the critical velocity for line emission, $\vkrit\approx25~\kms$ (similar for all lines; \citetalias{m22Schock}), which we estimate conservatively by setting $\mu=\muzpSch$ and $\zeta=\infty$ in Equation~(\ref{Gl:vthappGROSSzeta}). We saw in \citetalias{m22Schock} that the shock on the planet surface usually %
dominates the emission.

\section{Results}
 \label{Th:res}

We first show and validate detailed profiles for two parameter combinations (Section~\ref{Th:to}) and then apply our model to the population of forming planets (Section~\ref{Th:mPs}). Then, we compare the \PDS planets to our model (Section~\ref{Th:PDS}).

\subsection{Two examples}
 \label{Th:to}

\begin{figure*}[t]
 \centering
  \includegraphics[width=0.43\textwidth]{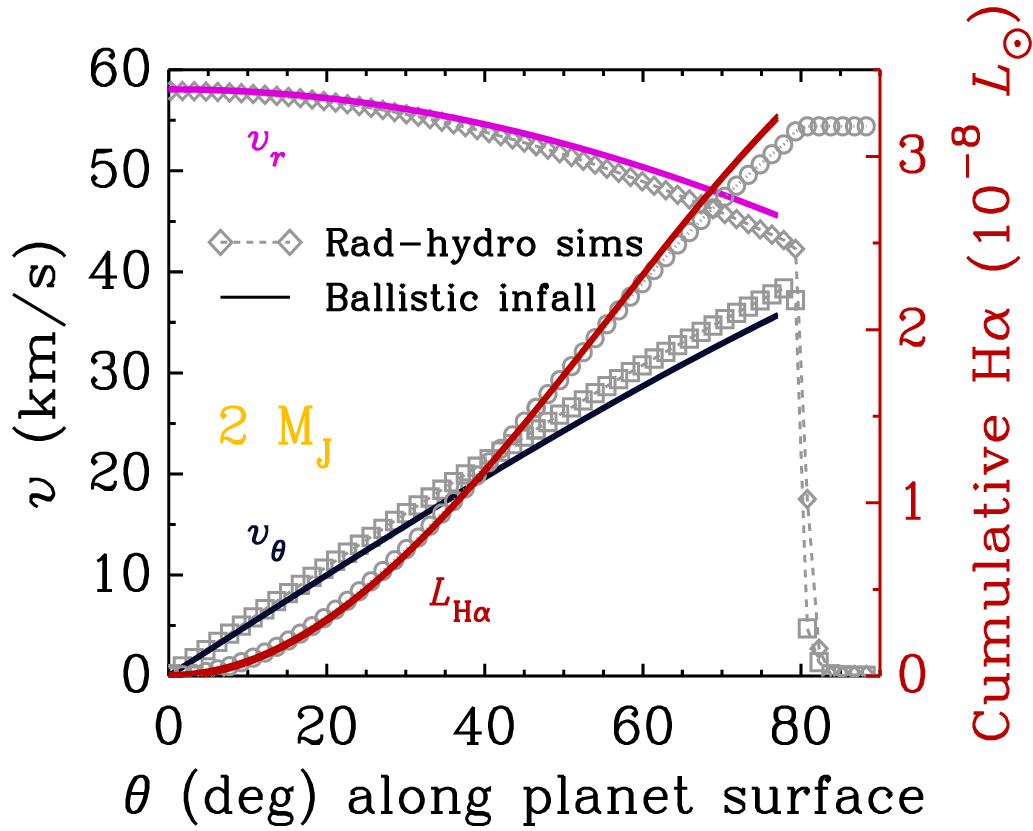}~
  \includegraphics[width=0.43\textwidth]{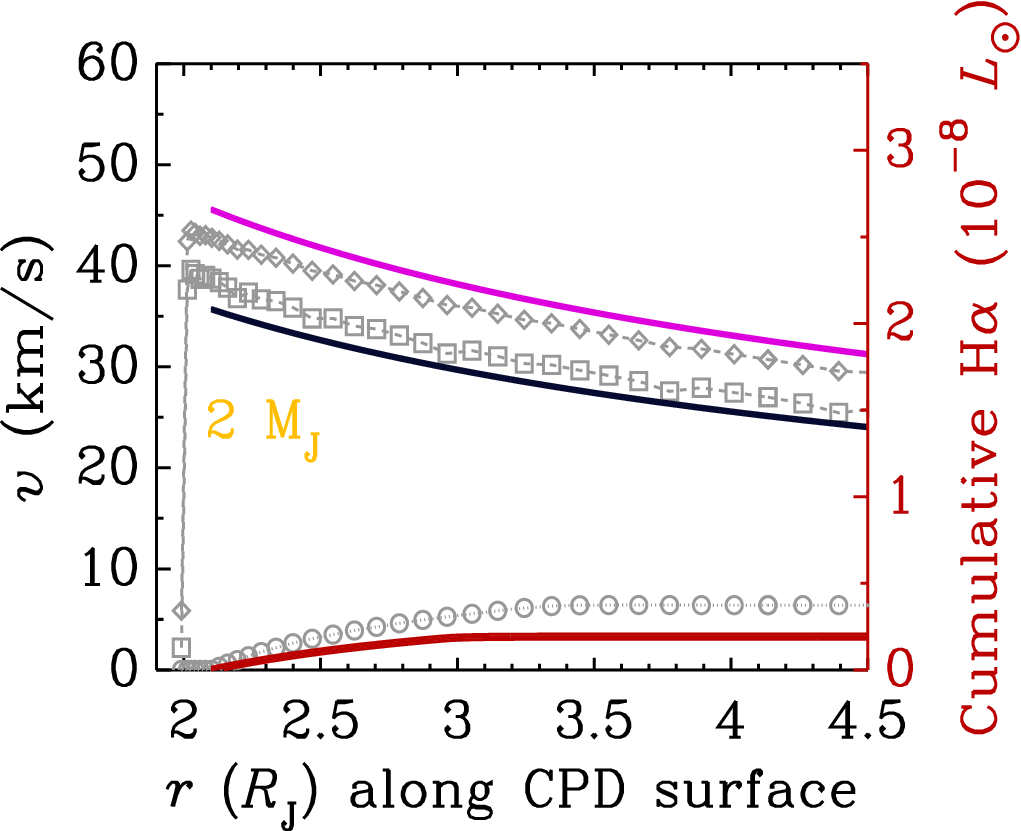}\\
  \includegraphics[width=0.43\textwidth]{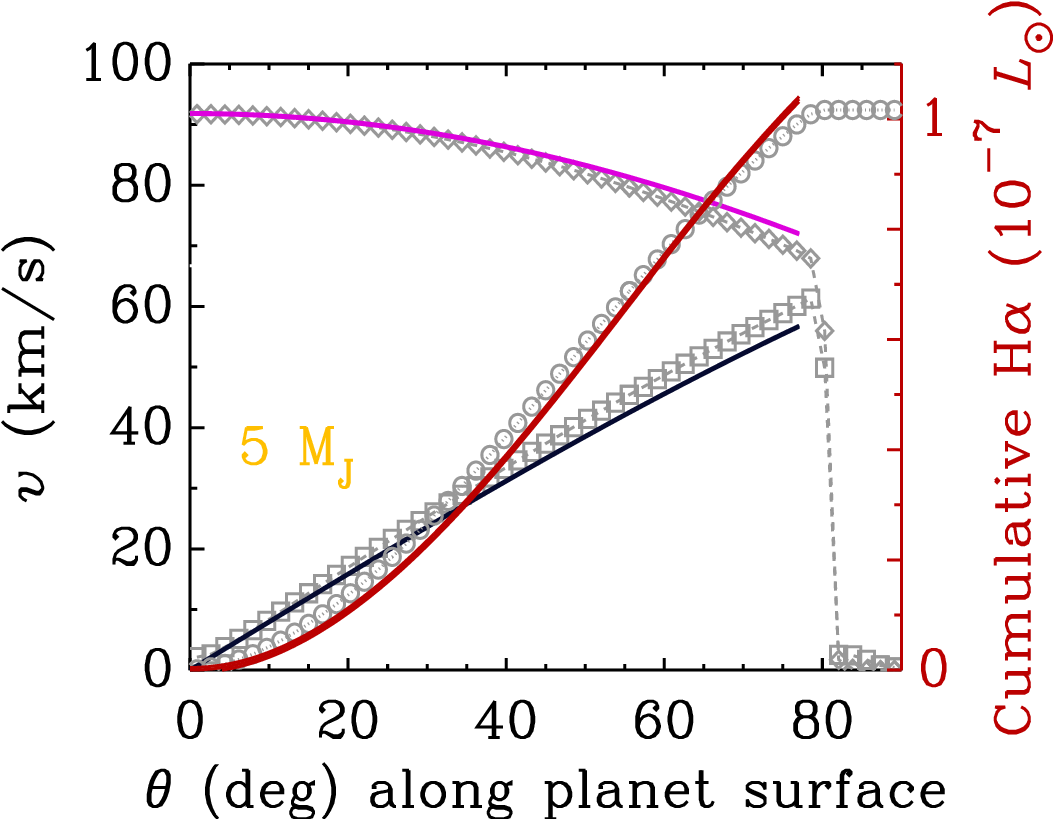}~
  \includegraphics[width=0.43\textwidth]{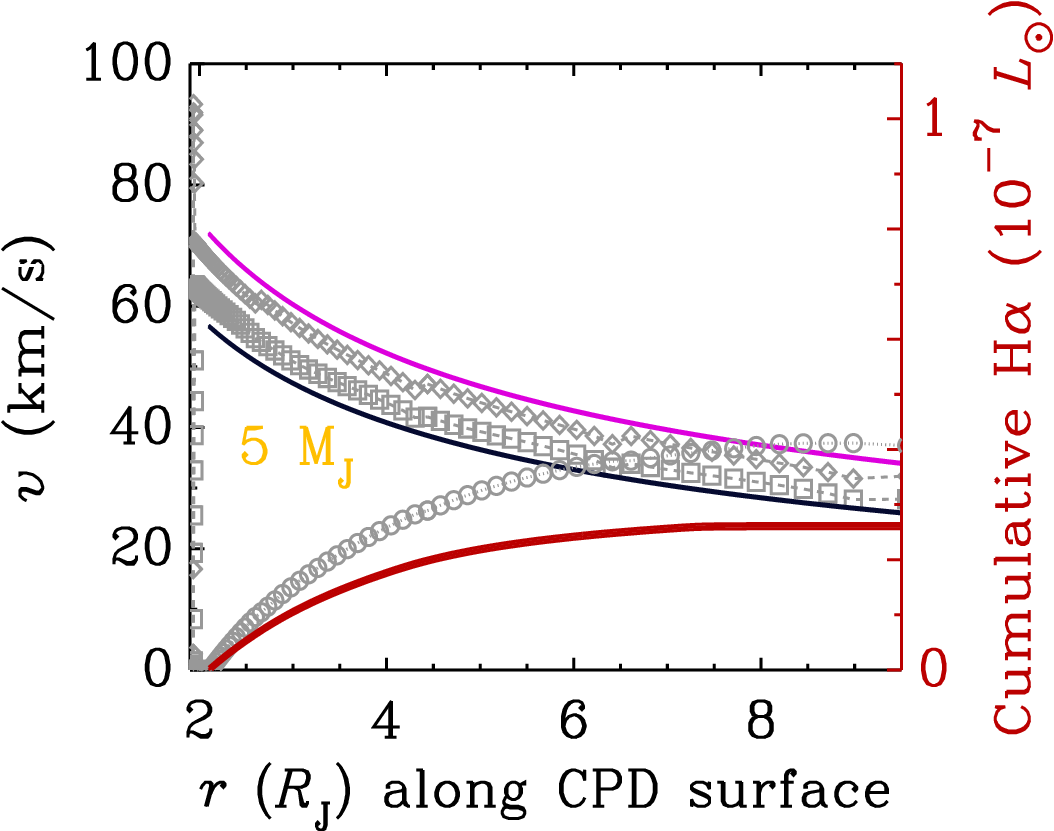}\\
\caption{%
Velocity of the preshock gas (left axes) and cumulative \Ha luminosity (right axes) for the shock at the surface at the planet (\textit{left panel}) and of the CPD (\textit{right})  %
for 2~\MJ (5~\MJ) top (bottom) row.
We set $\Sigma\approx0.02~\SigE$  %
and $\fzent\approx0.03$ to match the \MPktnettoHill and \MPktPdir of the detailed simulations (see Section~\ref{Th:fzent}).
With this, the ballistic model of this paper (coloured lines) reproduces excellently the radiation-hydrodynamical simulations of \citetalias{m22Schock} (grey points).
}
\label{Abb:vgl}
\end{figure*}

In Figure~\ref{Abb:vgl}, we show the velocities perpendicular to the shock \vVorSch on the planet surface and on the CPD surface.
We compare the ballistic solution to the simulations of \citetalias{m22Schock} for $\MP=2$~\MJ and 5~\MJ, using the $\Sigma$ and \fzent values calibrated to match the mass fluxes (Sections~\ref{Th:skalSigma} and~\ref{Th:fzent}). This involves only the kinetics and is independent of density.
There is a difference in the set-up: here, we assume zero-energy orbits with $v_\phi=0$ at $r_0$, whereas in \citetalias{m22Schock} we took the average shear flow into account. Nevertheless, at the shocks (on the planet surface and on the CPD), the agreement between the velocities from our model and from the simulations is excellent because the shock locations are deep in the potential, at $r\ll\RHill$.

As a litmus test of our model, we also show the \Ha fluxes, obtained with the microphysical model of \citet{aoyama18} in both cases but using the respective preshock velocities and densities.
The preshock densities \rhoVorSch (not shown) are nearly equal,
and despite the slight differences in \rhoVorSch and the respective \vVorSch, the local \Ha fluxes and their sums are very well reproduced,
even though $\FHa\sim\rhoVorSch\vVorSch^3$ (roughly; \citealp{aoyama18}) is a sensitive function of the preshock velocity.
Especially the main contribution, from the planet surface, is almost the same in the present model as in the simulations.
Therefore, our model is at least a reasonable approximation to the full simulations and can be applied to the population synthesis planets.

\subsection{With population synthesis}
 \label{Th:mPs} %

\begin{figure*}[t]
 \centering
  \includegraphics[width=0.62\textwidth]{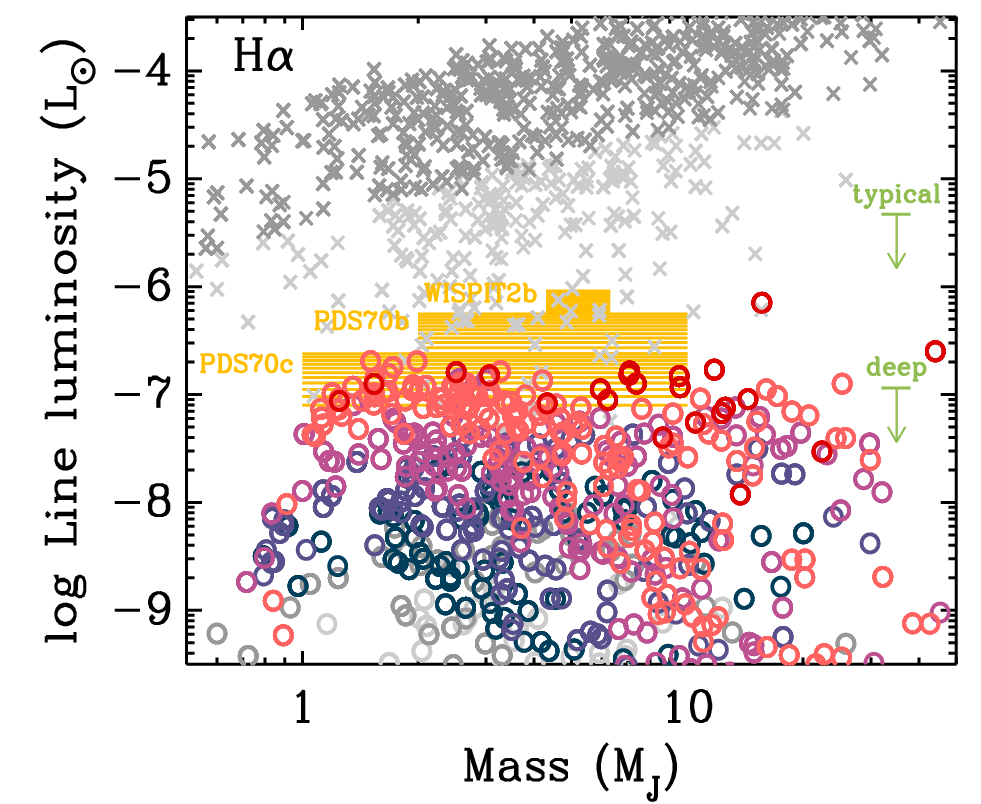}\\
  \includegraphics[width=0.32\textwidth]{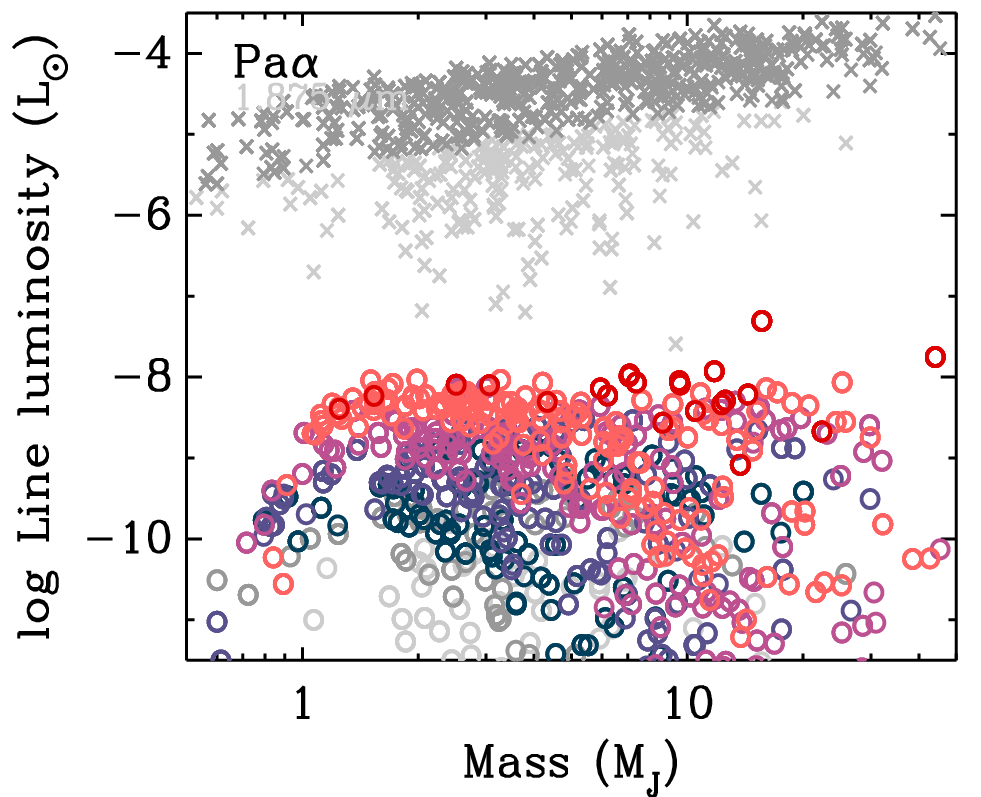}
  \includegraphics[width=0.32\textwidth]{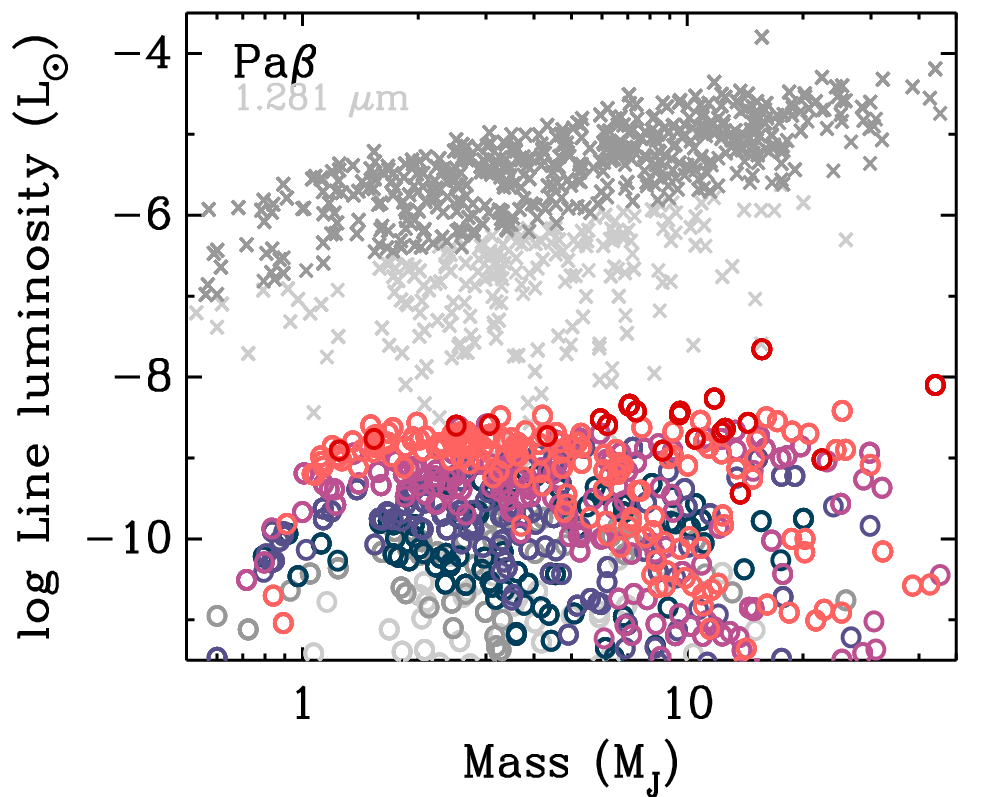}
  \includegraphics[width=0.32\textwidth]{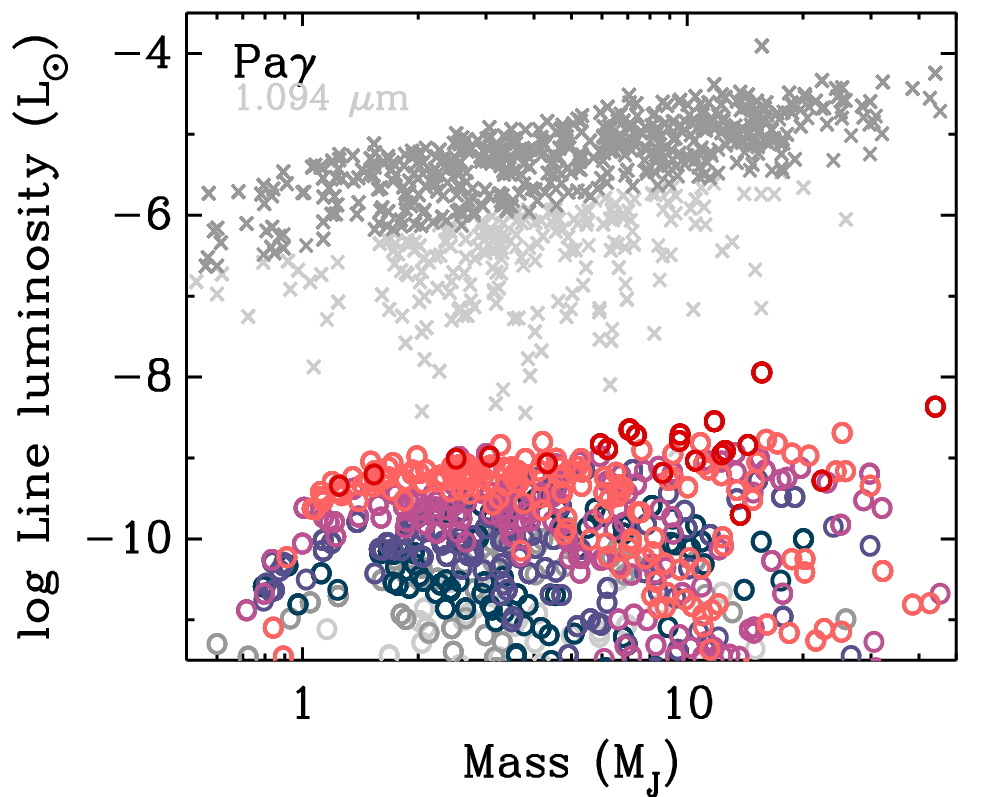}\\
  \includegraphics[width=0.32\textwidth]{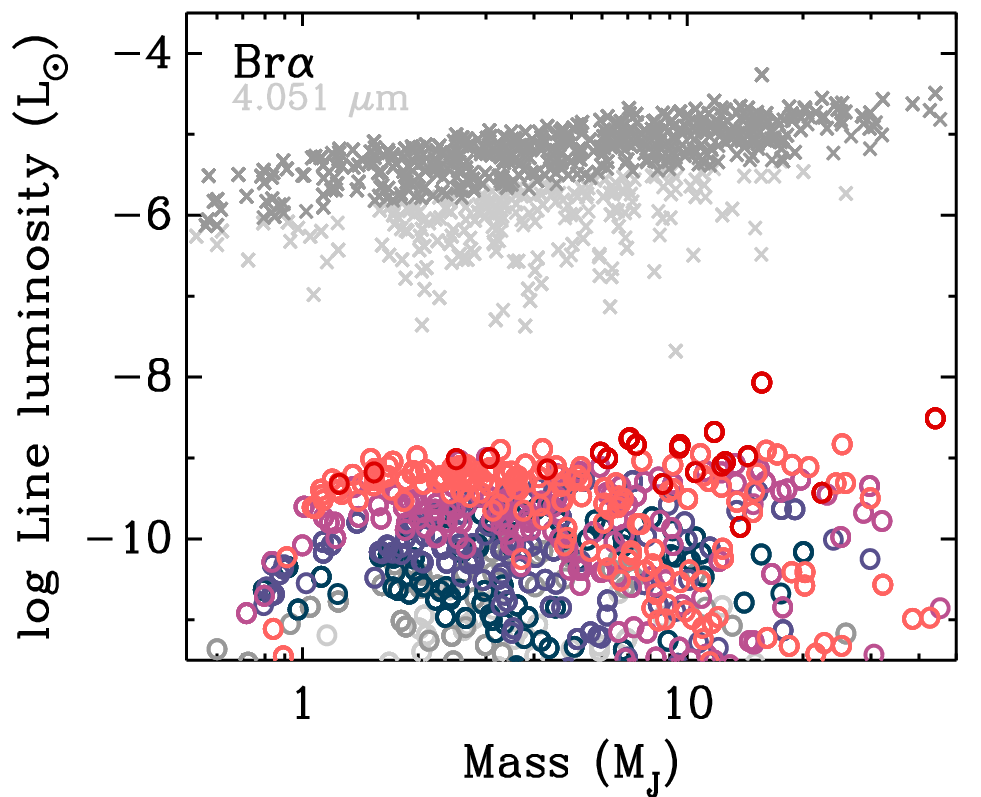}
  \includegraphics[width=0.32\textwidth]{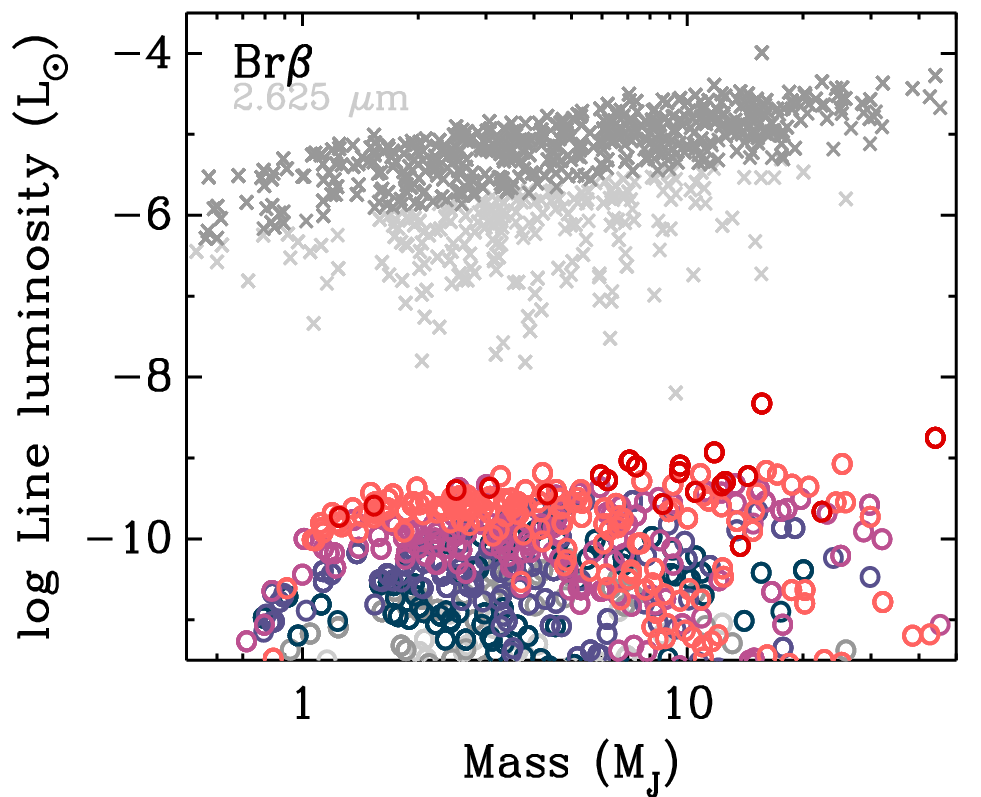}
  \includegraphics[width=0.32\textwidth]{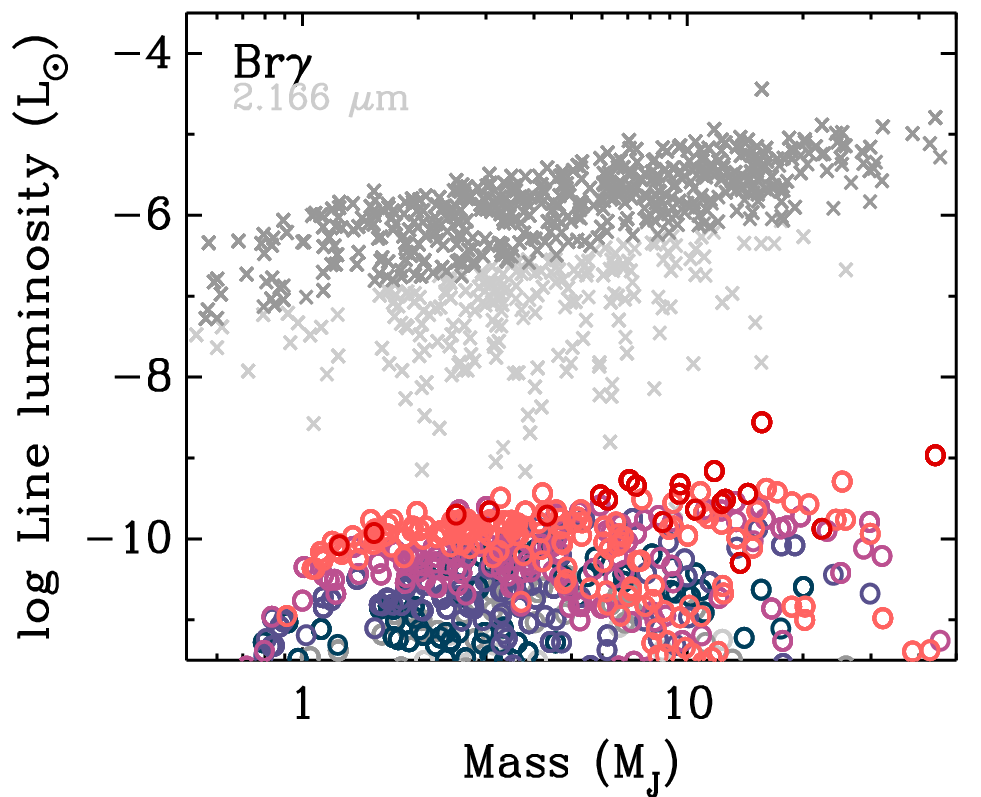}
\caption{%
Excess line luminosity
of accreting planets (open dots; see label in panels), summing the planet- and CPD-surface shocks.
We plot population synthesis planets %
during their main formation phase (1--5~Myr together).
Points are coloured by accretion rate as in Figure~\ref{Abb:Sigma}b.
The grey crosses (dark for $\MPktPopsynth\geqslant10^{-6}~\MPktEJ$, pale otherwise) are the %
estimate using extrapolated CTTS relationships  %
\citep{alcal17,Komarova+Fischer2020,rogers24}),
using the classical $\LAkk=G\MPktnettoHill\MP/\RP$.
\textit{Top panel}:
The non-dereddened luminosity and estimated masses of \PDSbc{} (see text)
are plotted for guidance (horizontal bars; height: variability of \LHa).
Also, WISPIT\,2\,b is shown \citep{close25b}.
Some synthetic planets are as faint as $\LHa\sim10^{-12}~\LSonne$ (not shown).
Segments on the right show typical or deep 99.99995-\%\ confidence-level upper limits, beyond 200~mas, from surveys or individual searches (respectively, \citealt{Cugno+2019,follette23,cugno23as209}), corresponding to $5\sigma$ for Gaussian residuals.
Limits for individual stars can easily differ by 1~dex.
}
\label{Abb:LHa+mehr}
\end{figure*}

In Figure~\ref{Abb:LHa+mehr}, we show one of the main results of this study: the line luminosities of an ensemble of forming planets.
The surface density at each planet is from Equation~(\ref{Gl:MPktvekstglychHill}), $\fzent=0.03$ from Section~\ref{Th:fzent} (but see   %
Section~\ref{Th:Scheibe}), and the aspect ratio $h$ is set to that of the PPD in the population synthesis at the position of the planet.
We examine \Ha and several hydrogen NIR lines accessible to ground- and space-based telescopes.

At \Ha, the maximum line-integrated luminosity is $\LHa\approx2\times10^{-7}~\LSonne$, with a large downward spread of values. The maximum luminosity is nearly independent of mass down to $\MP\approx1~\MJ$.
This mass independence might surprise given that \LHa is commonly thought to scale with the accretion luminosity \LAkk written as $\LAkk\approx G\MP\MPkt/\RP$; since \MPktvekst in the population synthesis does not depend strongly on mass and specifically does not decrease with mass (Figure~\ref{Abb:Sigma}a), one would expect that the maximum \LHa to grow with \MP, ignoring the variations in the planetary radius.

However, we assumed that \MPktvekst is the net Hill-sphere accretion rate \MPktnettoHill (Equation~(\ref{Gl:MPktvekstglychHill})), whereas the line-emitting accretion is closely related to \MPktPdir. In the expressions for \MPktPdir and \MPktnettoHill (Equations~(\ref{Gl:MPktPdirapprox}) and~(\ref{Gl:MPktnettoHillapprox})), two factors depend on \RHill: the crosssection and the density, which respectively increase and decrease with \RHill. The ratio $\RHill/\HPPPD=(\qth/3)^{1/3}$ is here sufficiently large ($\gtrsim1$) that the exponentially decreasing factor dominates.
We discussed this around Equation~(\ref{Gl:RHduerHPPPPPPa}), which was written for an example estimate of the temperature structure in the PPD.
In the population synthesis, the $\RHill/\HPPPD$ values even go up to $\approx6$ for closer-in 10-\MJ objects.  %
In Section~\ref{Th:MPktdirvglABT},
we find that the fraction of the mass flow reaching the planet directly drops from $f\sim10^{-2}$--$10^{-3}$ at $\MP\approx\MJ$ down to to $f\sim10^{-2}$--$10^{-7}$ at $\MP\sim10~\MJ$ (Figure~\ref{Abb:MPktPdir}b). This decrease of the line-emitting fraction with mass leads to the approximately constant upper envelope of luminosities.

\begin{figure*}[t]
 \centering
  \includegraphics[width=0.47\textwidth]{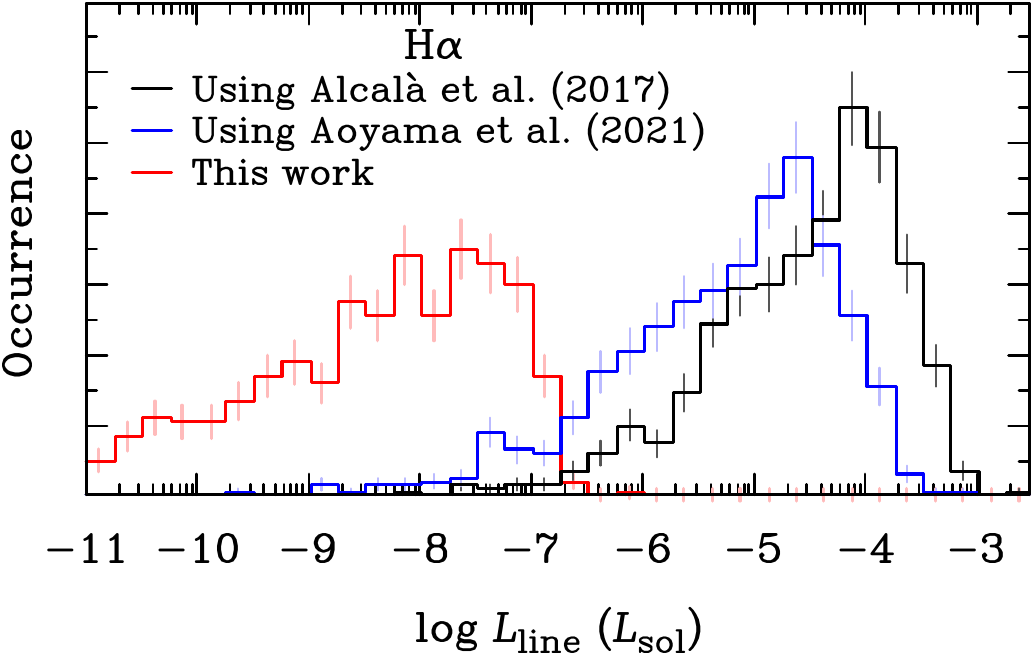}~~~~~
  \includegraphics[width=0.47\textwidth]{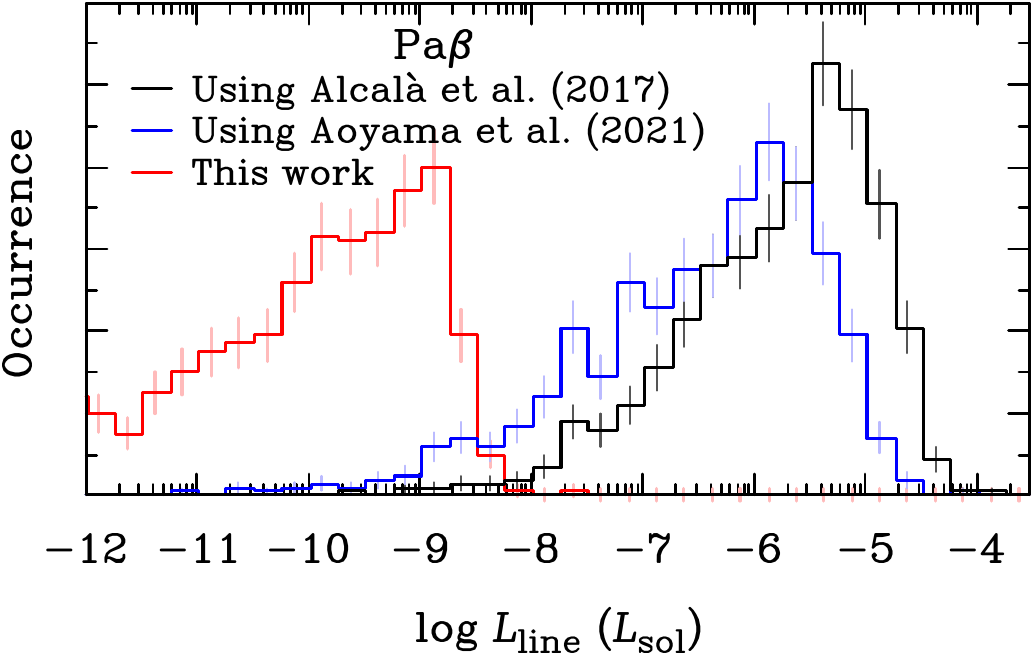}
\caption{%
Histograms of the luminosities from this work (red)
and when converting $\LAkk=G\MP\MPktnettoHill/\RP$ using
scalings for CTTSs extrapolated down to planetary masses (\citealt{alcal17}, consistent with the improved relations in \citealt{fiorellino25}; black)
or scalings for planetary-mass objects (\citealt{AMIM21L}; blue).
\textit{Left (right) panel}: for \Ha (\Pab).
}
\label{Abb:LLinieHist}
\end{figure*}

In the literature, including in works of which the present author is a main or co-author, a spherically-symmetric accretion geometry often has been implicitly or explicitly assumed when interpreting accretion-line fluxes. To explore the importance of taking at least approximately the 2D nature of accretion into account,
Figure~\ref{Abb:LHa+mehr} also shows the line luminosity predicted by scaling relationships for low-mass stars\footnote{The recent work of \citet{fiorellino25} updates and extends the scalings of \citet{alcal17} but they are essentially consistent.} \citep{alcal17,Komarova+Fischer2020} when extrapolated  %
to planetary masses.
As input into the relationships, we use $\LAkk=G\MP\MPktnettoHill/\RP$.
This is equivalent to Figure~7 in \citet{mordasini17} for \Ha, who used the relationship of \citet{rigliaco12} which, below $\LAkk\sim10^{-3}~\LSonne$, implies even higher luminosities (see comparison in Figure~1 of \citealt{AMIM21L}).
The resulting luminosities in \citet{mordasini17} reach up to $\LHa\approx10^{-3}~\LSonne$, which is three to four orders of magnitude higher than our predictions, and show a clear, linear mass dependence not seen in ours.
The line luminosity histograms in Figure~\ref{Abb:LLinieHist} make this even clearer: they are offset and the shapes of the distributions are different, with stronger peaks at high luminosities when using the CTTS relationships.
Therefore, using these leads to brightness predictions that might be overoptimistic, at least when using \MPktnettoHill for the \MPkt in \LAkk (see discussion in Section~\ref{Th:nondetect}).

In Figure~\ref{Abb:LHa+mehr} we also show the total luminosity for the first transitions of the Paschen and Brackett series, accessible with JWST. They have been detected at \Dlrmb \citep{betti22b,betti22c} and other low-mass objects such as \twb \citep{luhman23c,m23alois}.
As expected, the synthetic planets are fainter than at \Ha, with an upper envelope of luminosities that increases slightly towards high masses.
This trend is due to self-absorption in the postshock region becoming noticeable for the highest-mass planets at \Ha but not at the other lines.
Assessing the detectability is a separate undertaking
but we note that \twb has a \Pab line excess luminosity
$\LPab\approx4\times10^{-9}~\LSonne$ \citep{m23alois},
and except for the 2010 epoch (brighter by a factor of five),
GSC~06214-00210~b has $\LLinie\approx4\times10^{-8}~\LSonne$ on average \citep{demars23},
very similarly to \Dlrmb \citep{betti22b,betti22c}.
This suggests that for a sizeable fraction of the strongest accretors, their faintness compared to bolometric or \Ha-line luminosities should not hinder their observability.

By contrast, \citet{gangi22} fitted non-accretors at $\Teff\approx3300$--4500~K to have 
$\log \FPab < 4.85 + (\Teff - 3000)/2040$, %
where \FPab is the flux per emitting (and not detector) area
and quantities are in cgs units (\FEcgs and K).
If we extrapolate down this to $\Teff\approx2000$--3000~K appropriate for the two objects in \citet{demars23}, or $\Teff\approx1300$~K for \twb,
non-accretors would be predicted to have line luminosities below
$\LPab\approx(5$--$0.7)\times10^{-8}~\LSonne$, taking $\RP=2~\RJ$ as a typical radius for forming planets.

This limiting line luminosity is within a factor of ten higher than our estimates from the population synthesis, and comparable to that of the observed objects.
Both chromospheric activity and the accretion geometry at planetary masses are fraught with uncertainties, but chromospheric activity appears as the less secure process.
Therefore, we would lend less credence to it being the source of the observed line emission. As a corollary, this extrapolation of the \citet{gangi22} criterion (and similar ones at other lines) needs to be taken with caution because
it is likely pessimistic at these low masses.
More multiple-line detections as well as theoretical studies of the magnetic properties of young, low-mass accreting objects would be very useful.

\subsection[Comparison to the PDS 70 planets]{Comparison to the \PDS planets}
 \label{Th:PDS}

We compare the \Ha luminosities of the synthetic planets to the observed values of \PDSbc.
They are the first members of a possibly rapidly growing group (see three sentences on) of accreting planetary-mass companions found in a PPD, for which there could consequently
be a mass inflow from the Hill sphere. %
\citet{benisty21} found some evidence for this inflow at \PDSb, with the tentative detection of a ``streamer''.
Other low-mass companions (e.g., \Dlrmb; \citealp{eriksson20,ringqvist23,betti22b,betti22c}) are possibly too far away from the primary for its disc %
to be able to provide appreciable amounts of mass at that location ($\sim100$~au), and the modelled surface density at \twb \citep{ricci17,luhman23c} seems similarly too tenuous (at most $\Sigma\sim10^{-4}~\SigE$).
Other members of the group of low-mass accreting companions include the somewhat enigmatic AB~Aur~b \citep{currie22,zhou22,zhou23,currie24,biddle24}, which shows clear signs of accretion, possibly with an inverse P Cygni profile \citep{currie25b}, but whose signal might include contributions from scattered stellar light. Also, an \Ha-emitting source was found recently by \citet{li25} in the gap of the PPD around 2MASSJ16120668-3010270 (2M1612), which could very well be due to an accreting super-Jupiter, for which there are other lines of evidence \citep{sierra24a,ginski25}. A recent addition to the growing list is the 5-\MJ planet WISPIT\,2\,b \citep{vancapelleveen25a,close25b}, which has similarities with the \PDSbc planets in terms of mass, separation, and \Ha flux (see next paragraph), for instance. The field thus seems to be evolving quickly but for now, we will focus on the \PDS planets.  %

It has become clear that \PDSbc are variable on several timescales \citep{close25,zhou25}. The observed \Ha luminosities,
assuming isotropic emission and not correcting for possible extinction,
are around $\LHa\sim10^{-7}$~to $10^{-6}~\LSonne$, which
is up to a factor of several above the maximal \LHa from the population.
There are large uncertainties in the masses of \PDSbc,  %
with values $\MP\approx1$--$10~\MJ$ conceivable for both (\citealp{Haffert+2019,wang21vlti,hammond25,trevascus25}; see overview of the system properties in \citealt{shibaike24}). This uncertainty is
fortunately inconsequential for our comparison of the maximal expected line luminosity. However, due to the inclination of the system and the location of \PDSc close to the gap edge, \PDSc could be severely affected by extinction, so that in reality both planets could easily be intrinsically brighter than the population synthesis planets.

\begin{figure}[t]
 \centering
  \includegraphics[width=0.47\textwidth]{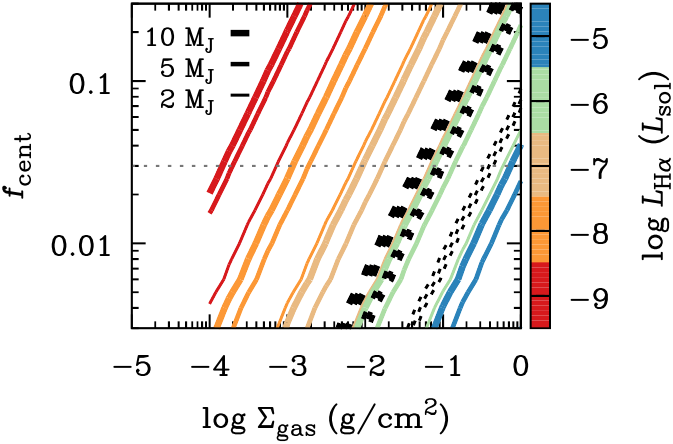}
\caption{%
Total \Ha luminosity (line colour) from the shocks on the planet and CPD surfaces
as a function of the local (gap-reduced) gas surface density $\Sigma$ (see text for discussion of meaning)
and the size of centrifugal radius relative to the Hill radius \fzent.
A thin dashed line highlights our fiducial value $\fzent=0.03$.
The radius is fixed to $\RP=2.1~\RJ$ but the mass is $\MP=2$, 5, 10~\MJ
(thin to thick lines). The black dashed lines are for $\LHa=(6.5\pm0.9)\times10^{-7}~\LSonne$ for \PDSb{} at one epoch \citep{zhou21} for each mass.
}
\label{Abb:LHa+mehrKont}
\end{figure}

To match the observed luminosity especially of \PDSb at least with the strongest accretors, an even smaller \fzent or a higher $\Sigma$ in the model would be required.
Figure~\ref{Abb:LHa+mehrKont} shows the combinations that work, assuming that the observed flux comes from the sum of the planet-surface and the CPD shocks.
For simplicity for this analysis, we use $\LHa=(6.5\pm0.9)\times10^{-7}~\LSonne$ for \PDSb from \citet{zhou21} but could repeat this with any of the other luminosities at different epochs.
We consider masses $\MP=2$--10~\MJ to fold in the mass uncertainty \citep{wang21vlti,hasegawa21}. We fix the radius to $\RP=2.1$~\RJ as a typical value but verified that it barely has an impact. Also, varying the CPD thickness (\muzpSch) does not matter. Finally, we set the PPD aspect ratio to $h=0.067$, following the modelling by \citet{bae19}.

For $\MP=2~\MJ$ and $\fzent=0.03$, Figure~\ref{Abb:LHa+mehrKont} shows that a surface density $\Sigma\approx0.4~\SigE$ in the gap at the location of \PDSb would be required to match the \Ha luminosity. A high planet mass $\MP=10~\MJ$ would require $\Sigma\approx0.05~\SigE$. Reducing \fzent by what is likely a generous factor of ten (i.e., $\fzent\sim0.0003$ is conservatively small) reduces the needed surface density to $\Sigma\sim0.005$--$0.05~\SigE$, whereas towards the canonical theoretical estimate $\fzent\approx1/3$, which is likely too high (see Section~\ref{Th:fzent} and references therein), the surface density tends towards $\Sigma\sim0.1$--1~\SigE.

All these gas surface density values are likely consistent with the detailed modelling effort of the \PDS system by \citet{pr23}. They derived 
azimuthal averages of
$\Sigma\approx0.003~\SigE$ at \PDSb and $\Sigma\approx0.007~\SigE$ at \PDSc.
While this is a factor $10$--100 times smaller than the surface densities we infer, simulations indicate that the local $\Sigma$ at the Hill sphere of the planet is also likely
at least ten times larger than the average in the gap
(see Figure~3 of \citealt{fung14}; Figure\footnote{Their figure shows the Hill-region-averaged $\Sigma$, not $\Sigma$ averaged at $1~\RHill$ around the planet, but yields a similar answer. The exact location in 3D simulations where the average should be calculated for our model is not well defined, so that an order of magnitude approximation seems acceptable.}~21 of \citealt{fung19}; J.~Bae 2023, private communication).
We recall also that in hydrodynamical simulations the gas entering the Hill sphere comes from a few Hill radii away \citep{tanigawa02}. In this case this may well still be in the gap, so that the argument holds, but in general, in our model the relevant density is the gas directly at the Hill sphere.
Nevertheless, the approximation of azimuthal symmetry around the planet makes the correct definition ambiguous.
Finally, even at surface densities up to $\Sigma\sim0.1~\SigE$, extinction from the disc should not be significant, as discussed above, at least at \PDSb.

\section{Discussion}
 \label{Th:Scheibe}  %

We discuss some aspects within or beyond our model.

\subsection{Interpreting the non-detections}
 \label{Th:nondetect}

Searches for accreting planets have revealed only few objects \citep{Cugno+2019,Zurlo+2020,xie20,hu22,follette23}
and our results suggest that an important factor is that only a small fraction of the accreting gas produces line emission.
Another factor is the lower line emission efficiency per unit accretion rate
at lower accretor masses \citep{AMIM21L}.
Together, this implies that extrapolated stellar relationships (e.g., \citealp{alcal17,fiorellino25}) overestimate survey yields, and that the non-detections are in fact less constraining than thought. This should be taken into account when interpreting survey yields in a statistical framework (e.g., \citealp{plunkett25}).

We did not analyse the separation dependence of the luminosity predictions since the migration of gas giants out to large separations is not very well understood, and eccentricity will modify the instantaneous planet--star separation too.
Even if there are planets around the stars surveyed so far and they are located in the favourable (background-limited) portion of detection curves, we do not expect that they would have been detected because they tend to be too faint for most instruments up to now.
Figure~\ref{Abb:LHa+mehr} suggests that %
the deep background-limited sensitivity values of \citet{cugno23as209} from MagAO-X---which benefits from major instrumental upgrades at MagAO---or the Hubble Space Telescope are only starting to scratch the tip of the iceberg. 
Recent discoveries with MagAO-X seem to corroborate this \citep{li25,close25b}.
Therefore, decreasing the inner working angle should give access to more planets \citep{nielsen19,close20}, but the more important factor might be an improvement in the sensitivity.

In the short term, using existing technology, it thus seems promising to survey a large number of stars while emphasising long integration times.  %
Even though there is some evidence that extinction plays a role in explaining the spectra of \PDSbc{} \citep{Stolker+20b,cugno21},  %
more recent data call this somewhat into question \citep{blakely25}. Also,
extinction by the accreting material close to the planet is predicted fortunately not to be a major factor overall\footnote{This was studied at \Ha. In the NIR, the dust opacity should be even smaller or at worse similar if it is grey instead of ISM-like \citep{cugno25}, but the gas opacity could be larger or smaller in narrow regions surrounding the other hydrogen emission lines.} \citep{maea21}.
Figure~\ref{Abb:LHa+mehr} shows that the maximal line luminosity is roughly constant with mass for all hydrogen lines. To first order, this implies that low-mass planets offer a similar contrast relative to their star compared to more massive ones and thus similar chances of detection.
However, more massive planets possibly open deeper gaps than lower-mass planets \citep[e.g.,][]{kanagawa17,zhang18}, which helps their detectability.
However, ultimately, it is not clear whether the planet-mass dependence of the remaining opacity in the gap is more important, and variability would likely complicate the matter.

\subsection{Limiting case for the accretion rate}

When using the accretion rates from the population synthesis, we made an important assumption: that the planet growth rate is equal to the net mass flux into the Hill sphere (Equation~(\ref{Gl:SigSkal})). If instead some of this mass flux served to let the CPD grow and the CPD were not efficient at transporting mass towards the planet,
a larger net mass flux into the Hill sphere would be required to have a given planet growth rate. This would imply an even higher surface density than we derived.
The accretion rate hitting the planet directly would be higher and also the line luminosities. Thus our current estimates provide lower limits on the surface densities.

We can turn this around and assume that the scenario we consider in this work---essentially, the consequence of angular momentum conservation, along with the assumption that there is no emission other than from the infall onto the planet and the CPD---describes the main factor for the small number of detected accreting planets. Then, the fact that the highest predicted luminosities are close to the detection limits up to now suggests that accreting planets cannot be much brighter, since otherwise more objects would have been detected. Therefore, not much less than \MPktnettoHill can be feeding the planets.
This could be used to put constraints on the timescale for radial mass transport in CPDs (\ab; \citealp{taylor24} %
since the viscous evolution of the disk needs to account for most of the mass gained by the planet, the direct infall rate being so small (see Section~\ref{Th:MPktdirvglABT} below). In that case, the planet gains mass almost only through boundary-layer accretion, which presumably does not lead to line emission (Section~\ref{Th:intro}).  %

\subsection{Heating of the planet by the shock}

\begin{figure}[t]
 \centering
  \includegraphics[width=0.45\textwidth]{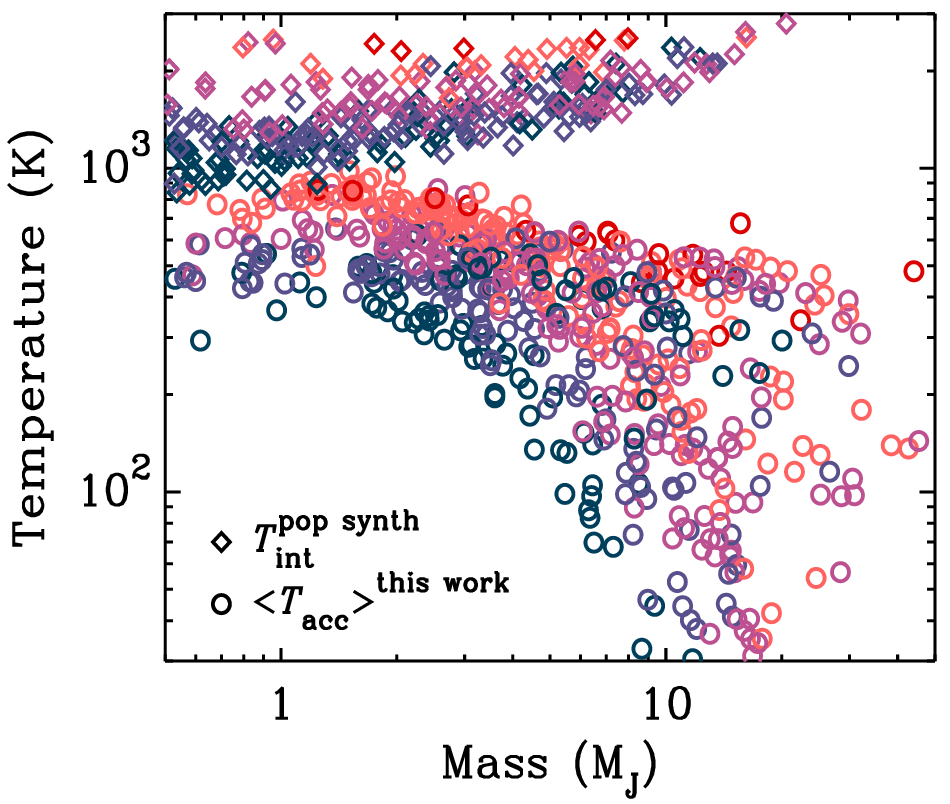}
\caption{%
Accretion temperature $\langle\TAkk\rangle$ as set by the rate \MPktPdir shocking on the surface of the planet (open circles),
compared to the interior temperature \Tint in the population synthesis (open diamonds).
Point colours show \MPktPopsynth as in the other plots but we omit the two weakest accretor bins in grey.
The \Tint of the strongest accretors could not be calculated (see text) but this leaves unchanged the conclusion that $\Tint\gg\langle\TAkk\rangle$ for each planet.
}
\label{Abb:TT}
\end{figure}

If we make the limiting assumption $\MPktvekst=\MPktnettoHill$, the growth rates in \citetalias{emsen21a} imply that only small to negligibly small rates $\MPktPdir\sim10^{-6}$--$10^{-9}$~\MPktEJ at $\MP=1~\MJ$ and $\MPktPdir\sim10^{-8}$--$10^{-14}~\MPktEJ$ at $\MP=10~\MJ$ reach the planets directly.
One can obtain the corresponding average accretion temperature from
$\LAkk\equiv4\pi\RP^2\sigSB\langle\TAkk\rangle^4$, setting $\LAkk=G\MP\MPktPdir/\RP$. This is a characteristic ``flux temperature'' on the accreting surface due to the shock. It ignores that a (small) equatorial band of the planetary surface is covered by the CPD.
Figure~\ref{Abb:TT} shows that $\langle\TAkk\rangle\approx400$--$1000$~K at $\MP\approx1~\MJ$, and $\langle\TAkk\rangle\lesssim500$~K, down to tens of kelvin, at $\MP\approx10~\MJ$. These temperatures correlate mostly with accretion rate $\MPktPopsynth=\MPktnettoHill$ up to roughly $\MP\approx3~\MJ$ but, at higher masses, decrease more strongly with increasing mass.

In spherical symmetry, we estimated in \citet{m18Schock} that essentially the entire \LAkk was released at the shock, but that the small fraction going inwards was much larger at least than in cold-start planets \citep{marl07}.
We make a similar approximate comparison by adding to Figure~\ref{Abb:TT} the intrinsic temperature \Tint of the ``cold-start'' population\footnote{As discussed in \citep{morda12_I}, the radii and luminosities of its synthetic planets are close enough to the ones of the ``hot-start population'' and thus of the \citetalias{emsen21a} we are using throughout this work. We cannot easily use \citetalias{emsen21a} here to study \Tint because the internal temperature is not available separately; only the summed contributions from the interior and the shock were saved.}  %
of \citet{morda12_I}.
We find that $\Tint\gtrsim1000$~K, increasing with accretion rate and mass.
For %
numerical reasons, we cannot extract the intrinsic temperature
for several synthetic planets, particularly towards high accretion rates.
However, this does not affect the finding that in all cases, \Tint is much larger than the accretion temperature for every planet.

Therefore, the gas shocking on the exposed planetary surface should not be able to heat the planet significantly. This contrasts with the spherically symmetric picture, in which the accreting gas brings in energy at a rate comparable to or larger than the intrinsic heat flux. The shock ram pressure is also likely much smaller than the photospheric pressure, so that the planet effectively does not notice the light sprinkling.

However, a complication in a multidimensional accretion geometry is the connection between the CPD and the planet.
For the stellar case, \citet{hartmann97} argue %
that neither magnetospheric accretion columns nor the boundary layer could transport much heat into the accretor.
However, it would be worthwhile to assess the importance of multidimensional effects, which does not seem to have been done even only for a comparable scenario. There are a few studies from the last decade that approach this topic but in every case, at least one aspect makes the work seem not directly applicable here: \citet{geroux16} assume uniform accretion over the whole surface into the accretor, \citet{hertfelder17} do not discuss the energy flux into the object and are in a likely physically very different regime in any case (for instance in terms of opacity, which sets the thermal time; they were studying white dwarfs), and \citet{dong21} do not include radiation transport. One could use a set-up as in \citetalias{m22Schock}, but put the inner radius of the computational domain further down, farther away from the planet radius (the shock location) to mitigate edge effects.
\subsection{Direct accretion rate}
 \label{Th:MPktdirvglABT}

One useful limit which \ab and \citet{taylor24} discussed is the case of no viscosity in the CPD and no magnetic field from the planet, so that the growth rate of the planet is set only by what falls directly onto it.
Equation~(49) of \citet{taylor24} provides
the resulting \MPktPdir in the thin-CPD ($\muzpSch\rightarrow0$) limit, that is, when the free planet surface reaches down to the midplane for the different inflow geometries:
\begin{equation}
 \label{Gl:MPktPdirABT}
  \MPktPdirABT = \fdir(\RP/\Rzent)\times\MPktnettoHill,
\end{equation}
where the subscript ``ABT'' refers to the series of papers by Adams, Batygin, and Taylor. The function $\fdir$ depends on the inflow function $f_i$. In the limit $\RP\ll\Rzent$, these functions \fdir are respectively
\begin{align}
 \label{Gl:fdir}
 \fdir =&\, (1.5,\; 1.0,\; 0.5,\; 0.42,\;0.38)\notag\\
       &\cdot(u_p,\;u_p,\;u_p,\;u_p^{1.5},\;u_p^2)\\
  u_p \equiv&\, \RP/\Rzent,\notag
\end{align}
for the five different geometries $f_i$, that is, 
a factor unity times $(\RP/\Rzent)^p$, where $p$ increases when going from ``Polar'' to 
``Equatorial''. From Figure~\ref{Abb:fi}, the case of a Gaussian density distribution, as we adopt here, can be seen as a ``super-equatorial'' scenario and would correspond to an even higher value of $p$.

Using Equations~(\ref{Gl:dmu0dmuapp^2}) and~(\ref{Gl:mu0min}), the fractional contribution to \MPktPdirABT (Equation~(\ref{Gl:MPktPdirABT})) by the streamlines landing between $\mu=\muzpSch$ and $\mu=0$ in the isotropic case studied by \citet{ab22} (which is the middle case ($i=3$) in the generalisation by \citet{taylor24} listed in Equation~(\ref{Gl:fdir})) is approximately $\Delta\mu_0/(1-\munaufp)\approx \left[\muzpSch\times\RP/(2\Rzent)\right]/\left[(1-\muzpSch)/2\right]\times\RP/\Rzent\approx\muzpSch/(1-\muzpSch)\approx0.3$ for $\muzpSch=0.23$.
In other words, in the isotropic case, the estimate with \MPktPdirABT (Equation~(\ref{Gl:MPktPdirABT})) is higher than \MPktPdir for a given density structure %
by tens of percent, which is negligible.

\begin{figure*}[t]
 \centering
  \includegraphics[width=0.43\textwidth]{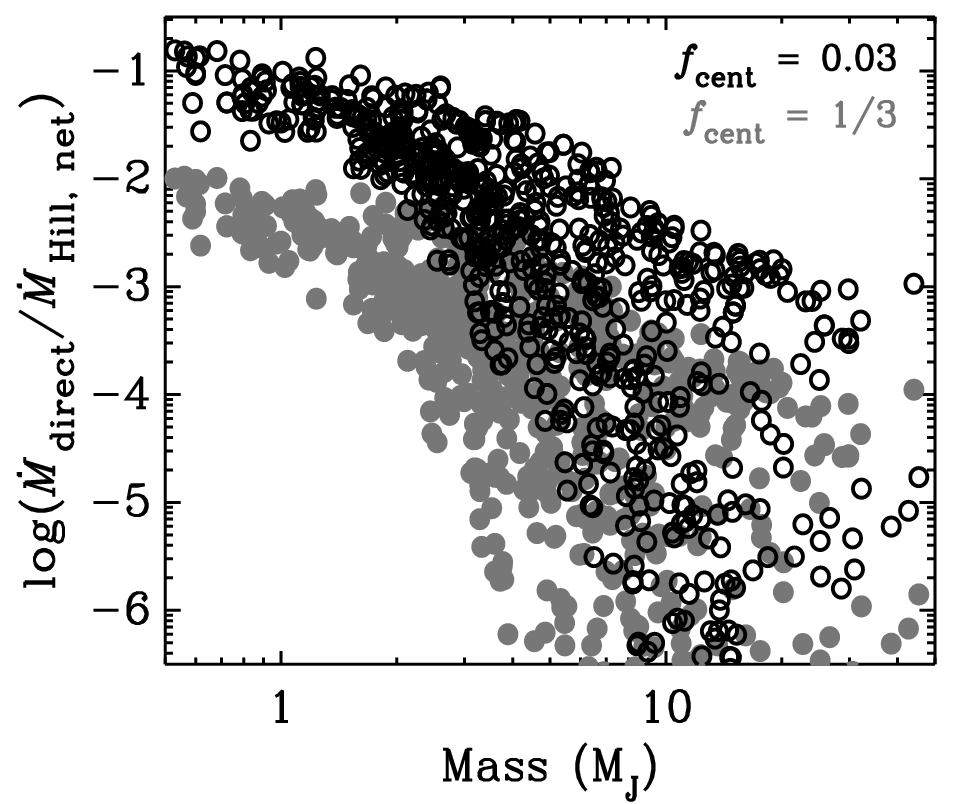}~~~~~~~%
  \includegraphics[width=0.43\textwidth]{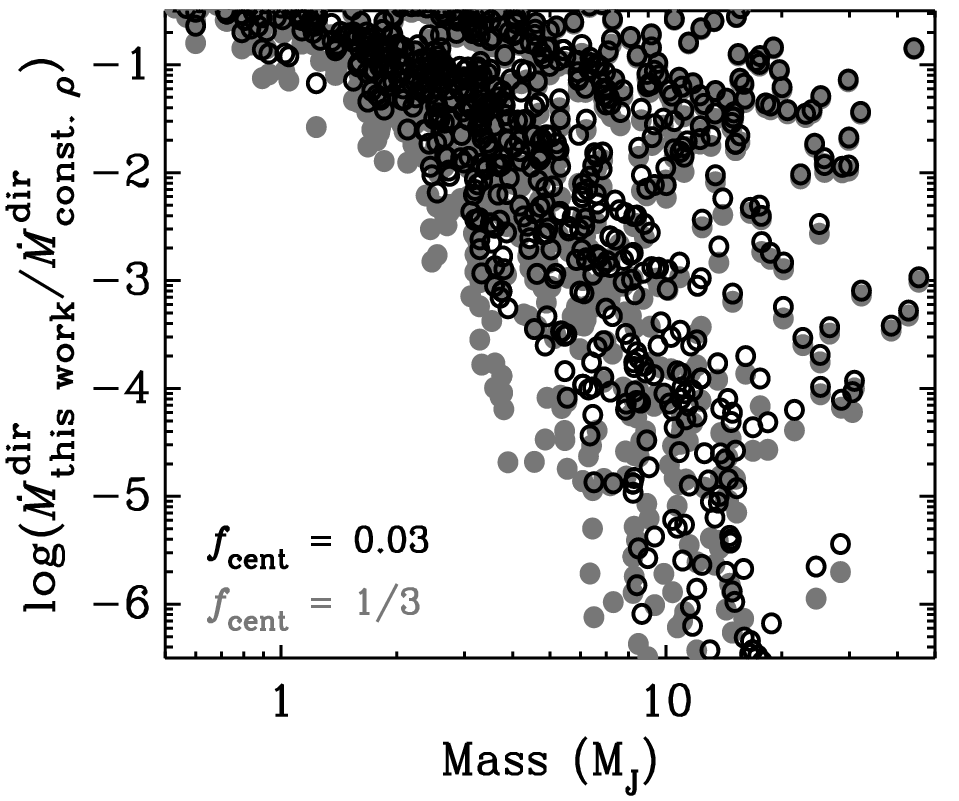} %
\caption{%
(\textit{Left panel}):
Fraction of \MPktnettoHill (integrating down to $\mu=\muzpSch$ by choice; see Equation~(\ref{Gl:MPktnettoHill})) that falls directly onto the planet (\MPktPdir). Black (grey) dots are for the assumption of $\fzent=0.03$ (1/3).
(\textit{Right panel}):
Comparison of the mass influx falling directly onto the planet when calculated in our set-up (Gaussian density profile at the outer edge and CPD-caused truncation of the equatorial inflow; Equation~(\ref{Gl:MPktPdirexakt}) or~(\ref{Gl:MPktPdirapprox})) relative to the result from a constant vertical density at the Hill radius integrated down to the equator (``Isotropic'' case in \citet{taylor24}).}
\label{Abb:MPktPdir}
\end{figure*}

Due to the presence of the planet, the true 3D density structure will not be an exact Gaussian set only by the stellar gravity (Equation~(\ref{Gl:rhor0})) and certainly not azimuthally symmetric around the planet, but inflow dominated by intermediate to low latitudes might be a reasonable qualitative set-up \citep[e.g.,][]{li23}. Nevertheless, the true angular dependence of the mass inflow is not known yet, and might vary with planet mass or accretion rate. Different works find different inflow geometries (Section~\ref{Th:accflow}).
Figure~\ref{Abb:MPktPdir}b %
compares the direct infall rate in our set-up to the ``Isotropic'' case of (Equation~(\ref{Gl:MPktPdirABT})).
As a consequence, at $\MP\approx1~\MJ$, the direct accretion rate is reduced by 2--10~times if one takes the density stratification into account, 
at 3~\MJ by a factor 2--100, and at 10~\MJ by up to more than 100,000. %

For completeness, in Figure~\ref{Abb:MPktPdir}b, we show the fraction of the net mass flux into the Hill sphere that reaches the planet directly because this is a useful proxy for the accretion-line-emitting mass rate (e.g., \Ha).
In \citetalias{m22Schock}, we confirmed by direct simulations that this is only a small fraction, on the order of 1\,\%.
Now more generally, from Equations~(\ref{Gl:MPktPdirapprox}) and~(\ref{Gl:MPktnettoHillapprox}), we have
\begin{align}
 \fdir = \frac{\MPktPdir}{\MPktnettoHill} = &\frac{1}{\sqrt{\pi}}\frac{\RP}{\fzent\RHill} \frac{1-\muzpSch}{\Derf}\left(\frac{\qth}{3}\right)^{1/3}\notag\\
 &\times\exp{\left[-\frac{1}{2}\left(\frac{\qth}{3}\right)^{-2/3}\right]}.
\end{align}

Within our framework here, we find that the conclusion of \citetalias{m22Schock} holds dramatically more towards large masses, where
\fdir
drops to $10^{-8}$--$10^{-3}$ at $\MP=10~\MJ$. %
The main driver of this decrease in $\fdir$ is not the increase of \RHill or $\RHill/\RP$ with \MP,  %
nor our choice of a fixed \fzent or \muzpSch. %
Equation~(\ref{Gl:MPktPdirapprox}) shows that the main factor is the exponentially dropping density (since $\RHill/\HPPPD$ is large enough), which comes from the hydrostatic structure of the PPD. What matters most for \MPktPdir is $\HPPPD/\RHill$, that is, the density near the pole $\rho_0(\RHill)$. Concerning the other factors,
the size of the polar cap \munaufp is of the same order for all planets, especially since we have fixed \fzent for all, and $\RHill^2\vFfinfty$ at \RHill scales as some order-(sub-)unity power of \MP.

The more relevant, exact quantity would be the line-flux-averaged mass flux relative to \MPktnettoHill: the emission from the planet surface is not uniform and the CPD close to the planet contributes a fraction which grows with increasing mass. Both of these facts are not reflected in \MPktPdir. However, given the approximations we have made, it might not be particularly enlightening to compute that average, and the correction relative to \MPktPdir\ would likely not be major.

\subsection{Neglect of magnetospheric accretion}
 \label{Th:keineMagAkk}

We discuss our neglect of 
the influence of magnetic fields on the accretion process onto the planet (magnetospheric accretion; e.g., \citealp{lovelace11,batygin18}) and therefore the line emission. A few comments can be made about this:
\begin{enumerate}
 \item There are theoretical reasons to expect young and therefore luminous gas giants to have $B\approx0.5$--2~kG magnetic fields, which might be strong enough to truncate the CPD \citep[e.g.,][]{katarzy16,hasegawa21}.
 \item It is not clear whether the gas is sufficiently ionised (thermally or not) to couple to the magnetic field, with non-ideal magnetohydrodynamical effects complicating the picture further \citep{batygin18,hasegawa21}.
 \item Empirically, a (perhaps growing) list of planetary-mass objects do seem to require magnetospheric accretion or be consistent with it to explain the shape or flux ratios of accretion-tracing hydrogen lines \citep{betti22b,betti22c,ringqvist23,demars23,m23alois,viswanath24,aoyama24twa,almendrosabad25,demars26}. 
 \item If planets accrete magnetospherically, the capture radius for the infalling gas would be larger than the planetary radius because gas falling within the apex of the magnetic fields would slide down towards the planetary surface (\citealp{batygin18}; Equation~(45) of \ab). This would increase by perhaps an order of magnitude the fraction of \MPktnettoHill landing indirectly on the surface of the planet (Equation~(45) of \citetalias{ab22} and Figure~1 of \citetalias{ab25}), a ``guaranteed'' minimum for the growth rate, independently of what happens to the gas which slides down the field lines outwards or lands directly on the CPD (Section~\ref{Th:MPkt}). However, it is not obvious how the gas capture affects the velocity components, especially whether the gas, when it reaches the footpoint of the field lines, still has or has regained an appreciable radial velocity component, which would let it emit significantly shock lines.
 \item In an elegant analysis, \citet{batygin25} posited that magnetospheric accretion was active during the formation of Jupiter and were able to infer from this detailed properties about Jupiter in an entirely self-consistent picture. This lends strong support to having $N\geqslant1$~gas giant in the Galaxy to which magnetospheric accretion applies.
\end{enumerate}
Therefore, magnetospheric accretion could be taking place at several accreting planets and this might change the detailed predictions of our model at an appreciable (factor-of-several) level. However, these modifications should be more quantitative than qualitative, and in any case, our model will apply---at least in principle---to those planets which are not accreting magnetospherically at a given time, if ever.

\section{Summary and conclusion}
 \label{Th:summconc}

In the context of only a few accreting planets having been discovered so far despite several searches, we studied how a non-spherically-symmetric accretion geometry onto a forming planet could help explain the rarity of accretion-line detections.
We developed a convenient approximation to the analytical multidimensional (2.5D) description of the gas flow onto an accreting gas giant forming in a PPD (\citealp{ulrich76}; \citealp{ab22}; \citealp{taylor24}; \citealp{ab25}). We extended this previous work by taking into account the exponential vertical stratification of the gas density at the Hill sphere and the possible cut-off of the equatorial inflow, due to the CPD. This set-up remains limited because we consider azimuthally symmetric polar-plane dynamics as opposed to a complex 3D geometry. However, this set-up does attempt to capture the important consideration of angular momentum conservation. Also, the true 3D flow does not seem known anyway at the smallest scales close to the planetary surface, due for example to the smoothing of the gravitational potential \citep{m22Schock}.

Especially for planets that are massive or forming closer in to their star, the non-negligible size of the Hill sphere compared to the pressure scale height of the PPD
dramatically reduces the fraction of the accretion rate hitting the planet surface directly.  %
We considered the limiting case that magnetospheric accretion does not occur, which should apply to some planets some of the time (Section~\ref{Th:keineMagAkk}).
Since line emission presumably requires a shock and scales strongly with the preshock velocity, only the planet surface and innermost regions of the CPD can emit. The planet still gains mass from the radially inward-directed flow in the CPD (see Equation~(A11) of \citealt{canupward02} or Equation~(89) of \abb), which in principle can be arbitrarily large and by definition needs to lead to a given planet mass within the lifetime of the CPD or the PPD. However, the actual transfer onto the planet occurs through a boundary layer, without a shock and therefore likely without line emission.
 most of this growth rate passes through a boundary layer that does not emit lines.

While the exact flow solution is tractable analytically (Appendix~\ref{Th:exakt}),
the small-polar-angle approximation we developed makes the solution easy to manipulate and to analyse, yielding insight into the parameter dependence.
Combining this with models of the hydrogen-line emission from the shocks on the planet and CPD \citep{aoyama18} led to predictions of the line luminosity of forming planets (Figure~\ref{Abb:LHa+mehr}).
We showed as main examples \Ha and bright NIR hydrogen lines, at several of which accreting low-mass objects have been detected.
More hydrogen lines are available upon request.
An important free parameter of the model is the centrifugal radius of the accretion flow, or equivalently its angular momentum bias $\ell$.
The classical value $\fzent=\Rzent/\RHill= 1/3$ (or $\ell=1$) is an upper bound
but, realistically, several effects reduce it (e.g., \ab; \abb).
Therefore, we calibrated it from 2.5D radiation-hydrodynamics simulations \citepalias{m22Schock} and found that $\fzent\approx0.03$ leads to a good match of the preshock properties of the gas. This agrees well with an independent line of reasoning (\abb). Future studies could benefit from a more detailed treatment, but our estimate leads to robustly conservative conclusions (Appendix~\ref{Th:fzent1/3}).

We applied our model to the distribution of 
properties of forming gas giants 
in the Bern model \citep{emsen21a}.
We found that \Ha line luminosities reach up to $\LLinie\approx2\times10^{-7}~\LSonne$ and that, remarkably, this is nearly independent of planet mass\footnote{This
  is either bad news---high-mass planets are as faint as low-mass planets---or good news---low-mass planets are as bright as high-mass ones. The latter should have more weight because observations and theory agree that low-mass gas giants are more numerous \citep[e.g.,][]{cumming08,nielsen19,emsen21b}.}.
These maximum values come from planets growing at a rate up to $\MPktvekst\sim10^{-5}~\MPktEJ$.
Peak luminosities for other lines are around $\LLinie\sim10^{-8}~\LSonne$ at \Paa, and $\LLinie\sim10^{-9}~\LSonne$ at \Pab, \Pag, \Bra, \Brg, or \Brg (roughly;  %
within a factor of three either way).

These maximum values and the \LLinie distributions are about ten times fainter than the \LAkk--\LLinie relationships of \citet{AMIM21L} or \citet{ma22}, and at least 1000~times fainter than what the extrapolation of \LAkk--\LLinie relationships for CTTSs (such as \citealt{alcal17} or \citealt{fiorellino25}; Figure~\ref{Abb:LLinieHist}) predicts.
The main reason is that angular momentum conservation lets only a minute fraction (of order one percent or orders of magnitude less; Figure~\ref{Abb:MPktPdir}b; \citetalias{m22Schock}) of the net mass flux into the Hill sphere emit accretion tracers such as \Ha.
Magnetospheric accretion, as for stars, would allow essentially the entire mass flux to emit.
Interestingly, the typical or even best-case sensitivities of \Ha surveys for accreting planets \citep{Cugno+2019,Zurlo+2020,xie20,hu22,follette23} are above the bulk of our predicted line luminosities. The same holds for the example of a deep search at AS~209 \citep{cugno23as209}.

An interesting recent development is the determination of $A_V\approx5$~mag in the wide gap around AS~209 \citep{cugno25}. While this value is considerable, our results suggest that, if planets were present in the surveyed systems, their non-detection would be due primarily to weak emission, and not to strong extinction (also not by the accretion flow itself; \citealt{maea21})
nor to particularly low accretion rates.
This is more of a broad-brush result, and bespoke studies of individual systems would be warranted and could lead to different conclusions.

Our semianalytical framework is meant more for population studies than for individual systems, due to the simplified set-up.
Nevertheless, we applied it to the measured \LHa of \PDSb %
to derive approximate constraints on the average gas surface density at the Hill sphere or over the Hill sphere (Section~\ref{Th:PDS}).
The constraints  %
seem consistent with the observations and modelling work of \citet{pr23}.
In this work, we did not include explicitly the possible implications of magnetospheric accretion on the line emission but discussed this in Section~\ref{Th:keineMagAkk}.
Planets accreting by magnetospheric accretion should be brighter for a given instantaneous mass growth rate because a larger fraction of this mass flux can generate lines: the gas joins the CPD at a large distance and therefore low velocity, moves radially inwards in the CPD, and is lifted by the magnetic field lines truncating the CPD and channelling the gas onto the surface of the planet, where it shocks. \Ha and other lines can be emitted in the accretion column, at the footpoint of the shock, or in both. There is a growing body of evidence that this mechanism can operate at planetary masses (see Section~\ref{Th:keineMagAkk}).

The take-away message from this work is that
accreting planets may %
be much fainter than usually thought %
for a given mass growth rate.
This is in fact a positive outlook because it implies that
the non-detections up to now
leave ample room for exciting discoveries.
It will be fruitful to push in closer to the star, where more forming planets should be found (e.g., \citealt{fernandes19,wittenmyer20,fulton21}, but see also \citealt{lagrange23}). Upcoming instruments such as the first-generation Midinfrared ELT Imager and Spectrograph (METIS) on the Extremely Large Telescope (ELT) \citep{ramsay18,feldt24} or RISTRETTO on the Very Large Telescope \citep{chazelas20,lovis24} should help see closer in. However, obtaining more \Ha photons with already-existing instrumentation is promising (for example the updated MagAO system at Magellan, MagAO-X; \citealp{males18,males24,close25b}).
Deeper integrations when targetting known young planets,
especially with high-resolution spectrographs, may reveal accretion tracers at objects down to barely a few Jupiter masses.

\begin{acknowledgments}
{\small
I am indebted to the referee for kind and impressively rapid reports with crucial comments on the tone of the text and for suggestions which improved the clarity of several points.
It is a pleasure to thank Yuhiko Aoyama very warmly for his model data,
Alexandre Emsenhuber for illuminating and inspiring discussions about accretion in the B\"arn model, and %
Christoph Mordasini for the secret correction factor needed for \citet{boden13}.
I thank Nick Choksi, Ruobing Dong, and Alexandros Ziampras for discussions about surface density.
I acknowledge the support of the DFG priority program SPP 1992 ``Exploring the Diversity of Extrasolar Planets'' (MA~9185/1), from the Swiss National Science Foundation under grant 200021\_204847 ``PlanetsInTime'',
and from the European Research Council (ERC) under the European Union's Horizon 2020 Research and Innovation Programme via the ERC Consolidator Grant ``PROTOPLANETS'' (Nr.~101002188; PI: M.~Benisty).
Parts of this work have been carried out within the framework of the NCCR PlanetS supported by the Swiss National Science Foundation.
This research has made use of the Astrophysics Data System Bibliographic Services of NASA.
No artificial intelligence (AI) was used in the course of this research.
Appendix~\ref{Th:exakt}, as noted there, used \texttt{Wolfram|alpha}, which clearly predates the current enthusiasm for AI, and does not involve AI in the traditional sense, as this engine is not ``mainly based on emulating human reasoning''
(see details at \url{https://www.wolframalpha.com/faqs}).
Most figures were produced using \href{https://github.com/gnudatalanguage/gdl}{\texttt{GDL}}, an actively-developed open-source drop-in alternative to \texttt{IDL}. The few other figures used \texttt{gnuplot} with the terminal \texttt{pdfcairo} and the font ``Hershey Complex'' thanks to \url{https://github.com/AstroJacobLi/smplotlib}, using font files made available in 2016 by Stewart C.\ Russell.  %
}
\end{acknowledgments}

\begin{appendix}

\section{Exact solution}
 \label{Th:exakt}

One can obtain the exact $\theta_0(\theta)$ and correspondingly also $d\mu_0/d\mu$ as a function of $\theta$. We give the expression for $\theta_0(\theta)$ here for completeness and plot them to verify the quality of our small-$\theta_0$ approximation. Comparisons of the flow lines are in Figure~\ref{Abb:flowducial} and~\ref{Abb:flowducialvarfzent}.

We solved Equation~(\ref{Gl:orb}) with the freely-available \texttt{Wolfram|alpha} Internet engine.
Apart from the trivial case $\zeta=0$, which yields $\mu=\mu_0$,
there are two different forms for the solution, depending on $\zeta$ and $\mu$.
The ``far-field'' form holds for any $\zeta\leqslant1$  %
and for $(\zeta<4,\mu>\mukrit)$ and is:
\begin{subequations}
\label{Gl:mu0exaktloin}
\begin{align}
 \mu_0 &= \frac{Q}{2^{1/3}\,3\zeta}+\frac{2^{1/3}\left(\zeta-1\right)}{Q},\\
 Q &\equiv 3\zeta^{2/3}\left(\mu+\sqrt{\mu^2-\frac{2^2}{3^3}\frac{\left(\zeta-1\right)^3}{\zeta}}\right)^{1/3}.
\end{align}
\end{subequations}
The ``near-field'' form applies for any $\zeta\geqslant4$
and for $(\zeta>1,\mu<\mukrit)$ and is:
\begin{subequations}
\label{Gl:mu0exaktproche}
\begin{align}
 \mu_0 &= 2\sqrt{\frac{\zeta-1}{3\zeta}}\cos\left[\frac{1}{3}\left(\tan^{-1} K\right)\right]\\
   K &\equiv \frac{1}{3^3\mu} \sqrt{2^2\,3^3\frac{\left(\zeta-1\right)^3}{\zeta}-3^6\mu^2}.
\end{align}
\end{subequations}
The transition happens at
\begin{equation}
 \mukrit(\zeta) = \sqrt{\frac{4}{27}\frac{\left(\zeta-1\right)^3}{\zeta}}
\end{equation}
For completeness, the corresponding critical $\mu_0$ is
\begin{equation}
 \munkrit(\zeta) = \left(\frac{\mukrit}{2\zeta}\right)^{1/3} + \frac{\zeta-1}{3\zeta}\left(\frac{2\zeta}{\mukrit}\right)^{1/3}.
\end{equation}

We briefly comment on how we obtained these expressions.
The real root of the cubic Equation~(\ref{Gl:orb}) has two terms, with $\sqrt{-1}$ respectively in the numerator and the denominator for the near-field case. The imaginary components however cancel for the real root.
Directly calling \texttt{Real[]} to extract  %
the real component of the root (both terms) did not yield a simple expression. Therefore, we applied \texttt{Real[]} for each term in turn and checked that the imaginary components are exactly the same but with opposite sign.
We found that the choice of the variable names given to \texttt{Wolfram|alpha} matters. To force being in one or the other regime to get the different solutions, we inserted non-rational constants as symbols ($\mu=1/e$ and $\zeta=\pi$ or $\zeta=\pi^\pi$); \texttt{Wolfram|alpha} kept them exactly (symbolically) but could determine in which regime a numerical evaluation would put the equation. By varying the assignments, we could probe the different regimes.

\begin{figure*}
 \centering
  \includegraphics[width=0.35\textwidth]{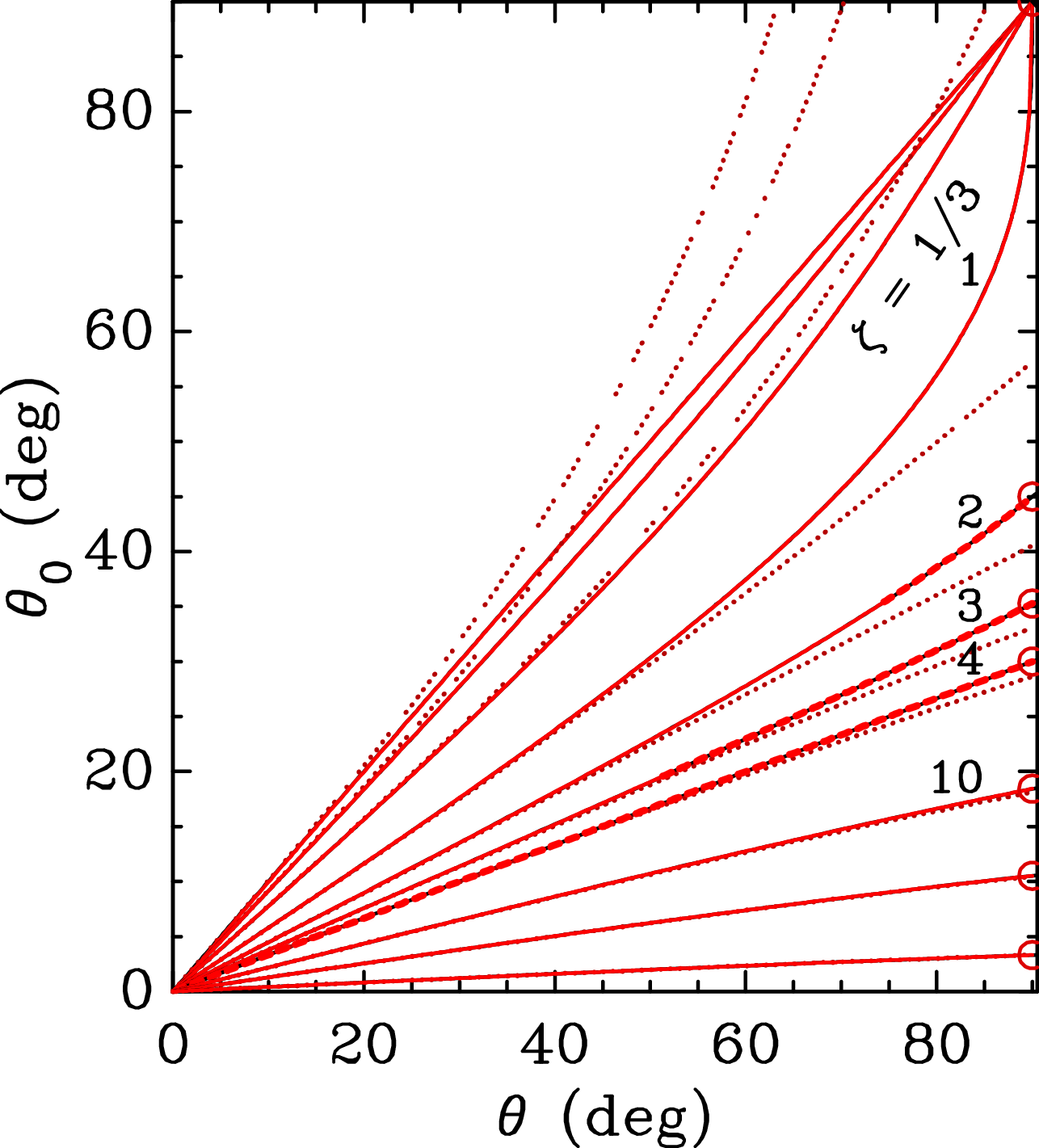}~~~~~~
  \includegraphics[width=0.362\textwidth]{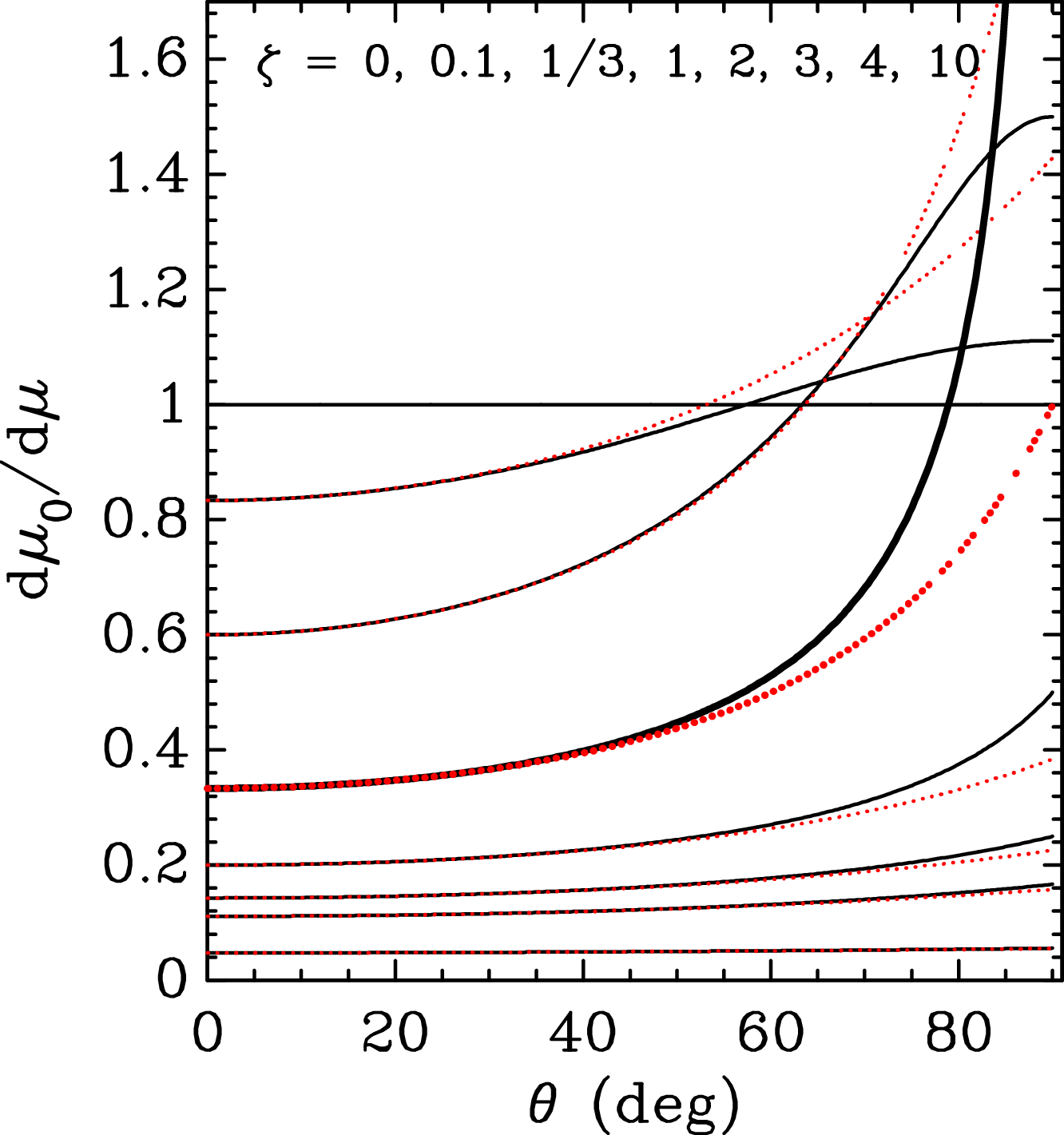}~~~
\caption{%
\textit{Left panel:}
Approximate starting angle (dotted red) compared to the exact solution (solid red) for different $\zeta$ values. The circles on the right show Equation~(\ref{Gl:mu0fuerMittelebene}).
\textit{Right panel:}
The quantity $d\mu_0/d\mu$ for $\zeta=0$, 0.1, 1/3, 2, 3, 4, 10 (top to bottom at the left $y$ axis). The approximation (red dotted curve; Equation~(\ref{Gl:dmu0dmuapp})) matches well, even towards large $\theta$. Only the $\zeta=1$ curve (bold) diverges towards the midplane, as it should (see left panel).  %
}
\label{Abb:thdm0}
\end{figure*}

In Figure~\ref{Abb:thdm0}, we show the exact trajectories and the second-order approximation. The match is excellent for small $\theta_0$.
We also compare the derivative $d\mu_0/d\mu$ needed for the density (Equation~(\ref{Gl:dmu0dmu}).
Finally, the streamline that lands on the midplane at an arbitrary
$r=\Rzent/\zeta$ is plotted as symbols in Figure~\ref{Abb:thdm0}a and is simply:
\begin{equation}
 \label{Gl:mu0fuerMittelebene}
\mu_0^{\rm mid} = \sqrt{ 1-\frac{1}{\zeta} }.
\end{equation}

We use the approximate expressions of Equation~(\ref{Gl:vKomponenteapprox}) but for completeness, we repeat the expressions for the exact velocity components, Equations~(9)--(11) of \ab{} (for $v_\theta$ and $v_\phi$, see also Equations~(28) and~(30) of \abb), but with our notation:
\begin{subequations}
\label{Gl:vKompexakt}
\begin{align}
 v_r &= -\vFfinfty \left[\frac{1}{2}\left(1-\frac{\mu}{\mu_0}\right)\right]^{1/2}\\  %
 v_\theta &= -\vFfinfty \left[ \frac{1-\mu_0^2}{1-\mu^2}(\mu_0^2-\mu^2)\frac{\zeta }{2}\right]^{1/2}\\
 v_\phi &= -\vFfinfty (1-\mu_0^2) \left[ \frac{\zeta}{2(1-\mu^2)} \right]^{1/2}.
\end{align}
\end{subequations}
Inserting Equations~(\ref{Gl:mu0exaktloin}) or~(\ref{Gl:mu0exaktproche}) into this yields the different velocity components at an arbitrary $(r,\theta)$ point.

\section{Assumption of a different centrifugal radius}
 \label{Th:fzent1/3}

\begin{figure}[t]
 \centering
  \includegraphics[width=0.43\textwidth]{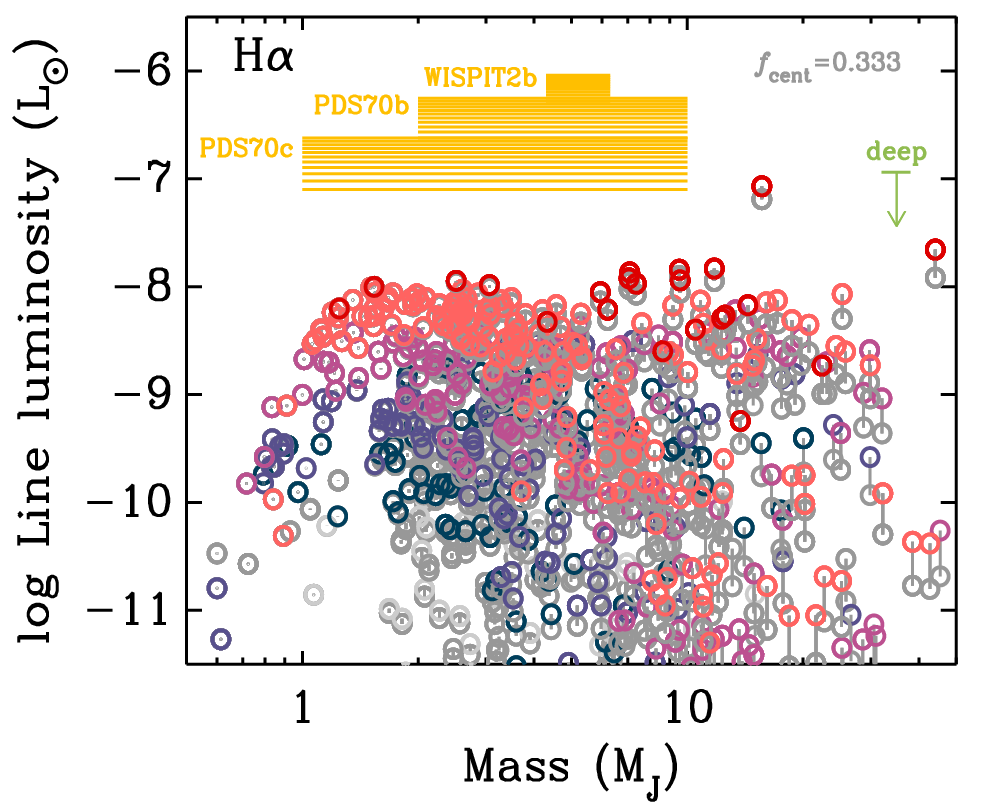}
\caption{%
As in Figure~\ref{Abb:LHa+mehr} but for $\fzent=1/3$ and only using our model. Each synthetic planet has two points, connected by a line, showing only the planet surface contribution \LPlObfl (lower, always grey) and the total $\LPlObfl+\LzpSch$ (higher point of the pair).
}
\label{Abb:LHa+mehr1/3}
\end{figure}

We show the total \Ha luminosity in Figure~\ref{Abb:LHa+mehr1/3} when assuming $\fzent=1/3$ for all planets because the main value we use, $\fzent=0.03$, was calibrated from only two simulations (for $\MP=2$~and 5~\MJ; Section~\ref{Th:fzent}).
The $\Sigma$ is the same as in Figure~\ref{Abb:LHa+mehr} since it is chosen to have $\MPktnettoHill=\MPktPopsynth$, which does not depend on \fzent, but now a smaller fraction of that mass influx reaches the planet directly or close to it. Most of it falls further out, making planets overall even less bright. Thus our main estimate is optimistic in this regard.

We also indicate (grey) the luminosity contribution coming only from the planet surface shock. It decreases with increasing planet mass but is at most only a factor of roughly two or three ($\approx0.3$--0.5~dex) smaller than the total.
The luminosity from the CPD depends on its thickness and its flaring. We assumed a relatively thin disc (angle from the midplane: $90\degr-\thzpSch\approx13\degr$) and no flaring. Thus, in reality, its contribution might be smaller than what is seen in Figures~\ref{Abb:LHa+mehr} and~\ref{Abb:LHa+mehr1/3}, especially towards lower planet masses if they have puffier CPDs \citep[e.g.,][]{krapp24,sagynbayeva25}. Nevertheless, we see that the minimum set by the planet surface is of the same order, so that our conclusions are robust.

\begin{figure}[t]
 \centering
  \includegraphics[height=0.27\textheight]{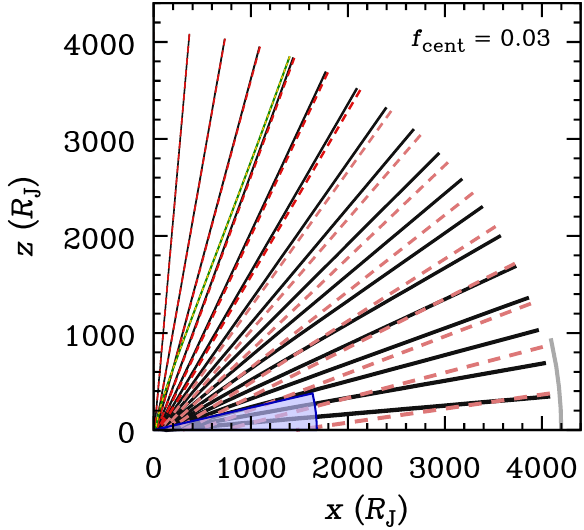}~~~
  \includegraphics[height=0.27\textheight]{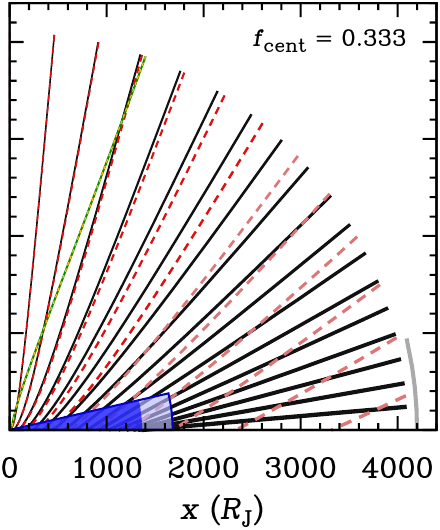}\\[0.7em]
  \includegraphics[height=0.27\textheight]{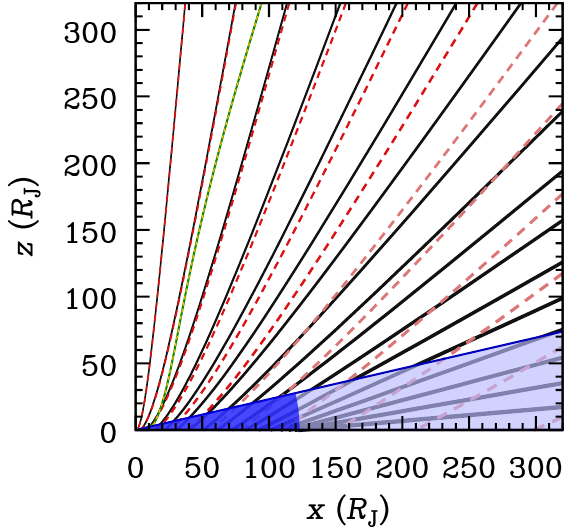}~~~
  \includegraphics[height=0.27\textheight]{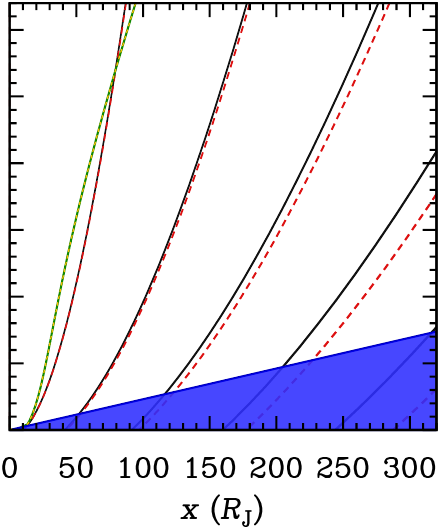}
  \caption{%
Flow at two zoom levels for the same parameters as in Figure~\ref{Abb:flowducial} (see description of elements there), but varying the relative centrifugal radius: $\fzent=0.03$ (\textit{left column}; as in Figure~\ref{Abb:flowducial}) and $\fzent=1/3$ (\textit{right}).
}
\label{Abb:flowducialvarfzent}
\end{figure}

For reference, we show in Figure~\ref{Abb:flowducialvarfzent} the flow pattern as in Figure~\ref{Abb:flowducial} but comparing $\fzent=0.03$ and 1/3. At a given scale, the small-$\theta_0$ approximation (Equation~\ref{Gl:muapprox}) is better for smaller \fzent values, which we use, but the flow is always qualitatively similar.

\end{appendix}

\bibliography{std}{}
\bibliographystyle{yahapj.bst}

\end{document}